\documentclass[11pt, a4paper]{article}
\usepackage{amssymb}
\usepackage{amsmath}
\usepackage{a4wide}
\usepackage{setspace}
\usepackage{makeidx}
\usepackage{natbib}
\usepackage{latexsym}
\usepackage{tikz}
\usetikzlibrary{decorations.markings}
\usepackage{mathpazo}
\usepackage[USenglish]{babel}
\usepackage{color}
\newtheorem{theorem}{Theorem}

\newtheorem{definition}{Definition}

\newtheorem{proposition}[theorem]{Proposition}

\newtheorem{lemma}[theorem]{Lemma}

\newtheorem{corollary}[theorem]{Corollary}

\def\D{\mathfrak{D}}
\def\B{\mathfrak{B}}

\def\F{\mathcal{F}}
\def\LP{\mathrm{LP}}
\def\GP{\mathrm{GP}}

\def\SED{\mathrm{SED}}
\def\GKL{\mathrm{GKL}}
\def\WCAP{\mathrm{WCAP}}
\def\WGCAP{\mathrm{WGCAP}}
\def\Fix{\mathrm{Fix}}
\def\Agg{\mathrm{Agg}}
\def\GAgg{\mathrm{GAgg}}

\begin{document}

\tikzset{->-/.style={decoration={
  markings,
  mark=at position #1 with {\arrow{>}}},postaction={decorate}}}

\title{Aggregating incoherent agents who disagree}
\author{Richard Pettigrew \\ Richard.Pettigrew@bris.ac.uk \footnote{I would like to thank Liam Kofi Bright, Remco Heesen, Ben Levinstein, Julia Staffel, Greg Wheeler, and two referees for \emph{Synthese} for helpful discussions that improved this paper.}\\ \\ forthcoming in \emph{Synthese}}
\date{\today}
\maketitle

Amira and Benito are experts in the epidemiology of influenza. Their expertise, therefore, covers a claim that interests us, namely, that the next `flu pandemic will occur in 2019. Call that proposition $X$ and its negation $\overline{X}$. Here are Amira's and Benito's credences or degrees of belief in that pair of propositions:\footnote{An agent's \emph{credence} in a proposition is her \emph{degree of belief} in it. That is, it measures how confident she is in the proposition. Sometimes these are called her \emph{subjective probabilities} or \emph{forecasts of probabilities}. I avoid the latter terminology since it might suggest to the reader that these credences are probabilistically coherent, and we are interested in this paper in cases in which they are not. As I write them, credences are real numbers in the unit interval $[0, 1]$. Others write them as percentages. Thus, where I write that Amira has credence 0.5 in $X$, others might write that she has credence 50\% in $X$ or that she is 50\% confident in $X$. Translating between the two is obviously straightforward.}
\begin{center}
\begin{tabular}{r|cc}
& $X$ & $\overline{X}$ \\
\hline 
Amira & 0.5 & 0.1 \\
Benito & 0.2 & 0.6
\end{tabular}
\end{center}
We would like to arrive at a single coherent pair of credences in $X$ and $\overline{X}$. Perhaps we wish to use these to set our own credences; or perhaps we wish to publish them in a report of the WHO as the collective view of expert epidemiologists; or perhaps we wish to use them in a decision-making process to determine how medical research funding should be allocated in 2018. Given their expertise, we would like to use Amira's and Benito's credences when we are coming up with ours. However, there are two problems. First, Amira and Benito disagree --- they assign different credences to $X$ and different credences to $\overline{X}$. Second, Amira and Benito are incoherent --- they each assign credences to $X$ and $\overline{X}$ that do not sum to 1. How, then, are we to proceed? There are natural ways to aggregate different credence functions; and there are natural ways to fix incoherent credence functions. Thus, we might fix Amira and Benito first and then aggregate the fixes; or we might aggregate their credences first and then fix up the aggregate. But what if these two disagree, as we will see they are sometimes wont to do? Which should we choose? To complicate matters further, there is a natural way to do both at once --- it makes credences coherent and aggregates them  all at the same time. What if this one-step procedure disagrees with one or other or both of the two-step procedures, fix-then-aggregate and aggregate-then-fix? In what follows, I explore when such disagreement arises and what the conditions are that guarantee that they will not. Then I will explain how these results may be used in philosophical arguments. I begin, however, with an overview of the paper. 

We begin, in section \ref{agg-cred}, by presenting the two most popular methods for aggregating credences: linear pooling ($\LP$) takes the aggregate of a set of credence functions to be their weighted arithmetic average, while geometric pooling ($\GP$) takes their weighted geometric average and then normalises that. Then, in section \ref{fix-cred} we describe a natural method for fixing incoherent credences: specify a measure of how far one credence function lies from the other, and fix an incoherent credence function by taking the coherent function that is closest to it according to that measure. We focus particularly on two of the most popular such measures: \emph{squared Euclidean distance} ($\SED$) and \emph{generalized Kullback-Leibler divergence} ($\GKL$). In section \ref{agg-fix-fix-agg}, we begin to see how the methods for fixing interact with the methods for aggregating: if we pair our measures of distance with our pooling methods carefully, they commute; otherwise, they do not. And we begin to see the central theme of the paper emerging: $\LP$ pairs naturally with $\SED$, while $\GP$  pairs with $\GKL$. In section \ref{agg-min-dist}, we note that, just as we can fix incoherent credence functions by minimizing distance from or to coherence, so we can aggregate credence functions by taking the aggregate to be the credence function that minimizes the weighted average distance from or to those credence functions. The aggregation methods that result don't necessarily result in coherent credence functions, however. To rectify this, in section \ref{agg-fix-together} we introduce the Weighted Coherent Aggregation Principle, which takes the aggregate to be the \emph{coherent} credence function that minimizes the weighted average distance from or to the credence functions to be aggregated. Up to this point, we have been talking generally about measures of the distance from one credence function to another, or only about or two favoured examples. In section \ref{bregman}, we introduce the class of additive Bregman divergences, which is the focus for the remainder of the paper. Our two favoured measures, $\SED$ and $\GKL$, belong to this class, as do many more besides. In section \ref{main-results} we come to the central results of the paper. They vindicate the earlier impression that linear pooling matches with squared Euclidean distance, while geometric pooling matches with generalized Kullback-Leibler divergence. Theorems \ref{sed-gkl-1-necessary} and \ref{sed-gkl-2-necessary} shows that the only methods of fixing or fixing-and-aggregating-together that commute with $\LP$ are those based on $\SED$, while the only methods that commute with $\GP$ are those based on $\GKL$. And Theorems \ref{d-agg-wcap}, \ref{agg-unique-1}, and \ref{agg-unique-2} describe the ways in which aggregating and fixing interact when we define both by minimizing distance. Finally, by section \ref{phil-sig}, we have all of our results in place and we can turn to their philosophical significance. I argue that these results can be used as philosophical booster rockets: on their own, they support no philosophical conclusion; but paired with an existing argument, either in favour of a way of aggregating or in favour of a particular measure of distance between credence functions, they can extend the conclusion of those arguments significantly. They say what measure of distance you should use if you wish to aggregate by $\LP$ or by $\GP$, for instance; and they say what aggregation method you should use if you favour $\SED$ or $\GKL$ over other measures of distance. In section \ref{partition}, we ask what happens when we lift a certain restriction that we have imposed throughout the earlier sections of the paper. And in section \ref{conclusion}, we conclude. The Appendix provides proofs for all of the results.

\section{Aggregating credences}\label{agg-cred}

Let's begin by considering two different methods for aggregating a collection of credence functions that are defined over the same set of propositions $\F = \{X_1, \ldots, X_m\}$. For the majority of the paper, we restrict attention to the case in which $\F = \{X_1, \ldots, X_m\}$ forms a partition.\footnote{That is, $X_1 \vee \ldots \vee X_m \equiv \top$ and $X_iX_j \equiv \bot$, for $1 \leq i \neq j \leq m$.} Then, in section \ref{partition}, we lift the restriction and see what happens. %Later, in section \ref{partition}, we'll ask what happens when we lift this restriction. The answer is that things get quite a bit more complicated! 

Let $C_\F$ be the set of credence functions over $\F$. And let $P_\F \subseteq C_\F$ be the set of coherent credence functions over $\F$.\footnote{Since $X_1, \ldots, X_m$ is a partition, $P_\F = \{c \in C_\F : c(X_1) + \ldots + c(X_m) = 1\}$.} Throughout, we take an agent's credence function to record her true credences. It doesn't record her reports of her credences, and it doesn't record the outcome of some particular method of measuring those credences. It records the credences themselves. Thus, we focus on cases in which our agent is genuinely incoherent, and not on cases in which they appear incoherent because of some flaw in our methods of measurement. 

An aggregation method is a function $T : (C_\F)^n \rightarrow C_\F$ that takes $n$ credence functions --- the agents --- and returns a single credence function --- the aggregate. Both aggregation methods we consider in this section appeal to a set of weights $\alpha_1, \ldots, \alpha_n$ for the agents, which we denote $\{\alpha\}$. We assume $\alpha_1, \ldots, \alpha_n \geq 0$ and $\sum^n_{k=1} \alpha_k = 1$.

First, \emph{linear pooling}. This says that we obtain the aggregate credence for a particular proposition $X_j$ in $\F$ by taking a weighted arithmetic average of the agents' credences in $X_j$; and we use the same weights for each proposition. The weighted arithmetic average of a sequence of numbers $b_1, \ldots, b_n$ given weights $\alpha_1, \ldots, \alpha_n$ is $\sum^n_{k=1} \alpha_kb_k = \alpha_1 b_1 + \ldots + \alpha_nb_n$. Thus:
\begin{quote}
\textbf{Linear Pooling (LP)} Let $\{\alpha\}$ be a set of weights. Then
\[
\LP^{\{\alpha\}}(c_1, \ldots, c_n)(X_j) = \alpha_1 c_1(X_j) + \ldots + \alpha_nc_n(X_j) = \sum^n_{k=1} \alpha_kc_k(X_j)
\]
for each $X_j$ in $\F$.
\end{quote}
Thus, to aggregate Amira's and Benito's credences in this way, we first pick a weight $0 \leq \alpha \leq 1$. Then  the aggregate credence in $X$ is $0.5 \alpha + 0.2 (1-\alpha)$, while the aggregate credence in $\overline{X}$ is $0.1\alpha + 0.6 (1-\alpha)$. Thus, if $\alpha = 0.4$, the aggregate credence in $X$ is $0.32$, while the aggregate credence in $\overline{X}$ is $0.4$. (See Figure \ref{amira-benito-sed} for an illustration of the effect of linear pooling on Amira's and Benito's credences.) Notice that, just as the two agents are incoherent, so is the aggregate. This is typically the case, though not universally, when we use linear pooling. 

Second, we consider \emph{geometric pooling}. This uses weighted \emph{geometric} averages where linear pooling uses weighted \emph{arithmetic} averages.  The weighted geometric average is $\prod^n_{i=1} b^{\alpha_i}_i = b_1^{\alpha_1} \times \ldots \times b_n^{\alpha_n}$. Now, when all of the agents' credence functions are coherent, so is the credence function that results from taking weighted \emph{arithmetic} averages of the credences they assign. That is, if $c_k(X_1) + \ldots + c_k(X_m) = 1$ for all $1 \leq k \leq n$, then
\[
\sum^n_{k=1} \alpha_kc_k(X_1) + \ldots + \sum^n_{k=1} \alpha_kc_k(X_m) = 1
\]
However, the same is not true of weighted \emph{geometric} averaging. Even if $c_k(X_1) + \ldots + c_k(X_m) = 1$ for all $1 \leq k \leq n$, there is no guarantee that
\[
\prod^n_{k=1} c_k(X_1)^{\alpha_k} + \ldots + \prod^n_{k=1} c_k(X_m)^{\alpha_k} = 1
\]
Thus, in geometric pooling, after taking the weighed geometric average, we need to normalize. So, for each cell $X_j$ of our partition, we first take the weighted geometric average of the agents' credences $c_k(X_j)$, and then we normalize the results. So the aggregated credence for $X_j$ is
\[
\frac{\prod^n_{k=1} c_k(X_j)^{\alpha_k}}{\prod^n_{k=1} c_k(X_1)^{\alpha_k} + \ldots + \prod^n_{k=1} c_k(X_m)^{\alpha_k}} = \frac{\prod^n_{k=1} c_k(X_j)^{\alpha_k}}{\sum^m_{i=1} \prod^n_{k=1} c_k(X_i)^{\alpha_k}} 
\]
That is,
%However, if we do this separately for different partitions, the results may not cohere with one another. Thus, we pick the most fine-grained partition in $\F$, which we call $\W$ and which we might think of as specifying the set of possible worlds grained as finely as $\F$ will allow. Then:
\begin{quote}
\textbf{Geometric Pooling (GP)} Let $\{\alpha\}$ be a set of weights. Then
\[
\GP^{\{\alpha\}}(c_1, \ldots, c_n)(X_j) = \frac{\prod^n_{k=1} c_k(X_j)^{\alpha_k}}{\sum^m_{i=1} \prod^n_{k=1} c_k(X_i)^{\alpha_k}} 
\]
for each  $X_j$ in $\F$.
\end{quote}
Thus, to aggregate Amira's and Benito's credences in this way, we first pick a weight $\alpha$. Then  the aggregate credence in $X$ is $\frac{0.5^\alpha  0.2^{1-\alpha}}{0.5^\alpha  0.2^{1-\alpha} + 0.1^\alpha  0.6^{1-\alpha}}$, while the aggregate credence in $\overline{X}$ is $\frac{0.1^\alpha + 0.6^{1-\alpha}}{0.5^\alpha  0.2^{1-\alpha} + 0.1^\alpha  0.6^{1-\alpha}}$. Thus, if $\alpha = 0.4$, the aggregate credence in $X$ is $0.496$, while the aggregate credence in $\overline{X}$ is $0.504$. (Again, see Figure \ref{amira-benito-gkl} for an illustration.) Note that, this time, the aggregate is guaranteed to be coherent, even though the agents are incoherent.

\section{Fixing incoherent credences}\label{fix-cred}

Amira has incoherent credences. How are we to fix her up so that she is coherent? And Benito? In general, how do we fix up an incoherent credence function so that it is coherent? A natural thought is that we should pick the credence function that is as similar as possible to the incoherent credence function whilst being coherent --- we might think of this as a method of minimal mutilation.\footnote{Such a fixing procedure is at least suggested by the second central result of \citep[204]{debona2016gi}. We will meet the principle of minimal mutilation again in section \ref{phil-sig}.}

For this purpose, we need a measure of distance between credence functions. In fact, since the measures we will use do not have the properties that mathematicians usually require of distances --- they aren't typically \emph{metrics} --- we will follow the statisticians in calling them \emph{divergences} instead.  A divergence is a function $\D : C_\F \times C_\F \rightarrow [0, \infty]$ such that (i) $\D(c, c) = 0$ for all $c$, and (ii) $\D(c, c') > 0$ for all $c \neq c'$. We do not require that $\D$ is symmetric: that is, we do not assume $\D(c, c') = \D(c', c)$ for all $c, c'$. Nor do we require that $\D$ satisfies the triangle inequality: that is, we do not assume $\D(c, c'') \leq \D(c, c') + \D(c', c'')$ for all $c, c', c''$. Now, suppose $\D$ is a divergence. Then the suggestion is this: given a credence function $c$, we fix it by taking the coherent credence function $c^*$ such that $\D(c^*, c)$ is minimal; or perhaps the coherent credence function $c^*$ such that $\D(c, c^*)$ is minimal --- since $\D$ may not be symmetric, these two ways of fixing $c$ might give different results. Thus:\footnote{Recall: $P_\F$ is the set of coherent credence functions over $\F$.}
\begin{quote}
\textbf{Fixing }  Given a credence function $c$, let
\[
\Fix_{\D_1}(c) = \underset{c' \in P_\F}{\mathrm{arg\, min}\ } \D(c', c)
\]
and
\[
\Fix_{\D_2}(c) = \underset{c' \in P_\F}{\mathrm{arg\, min}\ } \D(c, c')
\]
\end{quote}
Throughout this paper, we will be concerned particularly with fixing incoherent credence functions using the so-called \emph{additive Bregman divergences} \citep{bregman1967rm}. I'll introduce these properly in section \ref{bregman}, but let's meet two of the most famous Bregman divergences now:
\begin{quote}
\textbf{Squared Euclidean Distance (SED) }
\[
\SED(c, c') = \sum_{X \in \F} (c(X) - c'(X))^2
\]
\end{quote}
This is the divergence used in the {least squares method} in data fitting, where we wish to measure how far a putative fit to the data, $c$, lies from the data itself $c'$. For arguments in its favour, see \citep{selten1998acq, leitgeb2010ojb1, dagostino2010easp, pettigrew2016alc}.
\begin{quote}
\textbf{Generalized Kullback-Leibler (GKL) } 
\[
\GKL(c, c') = \sum_{X \in \F} \left  (  c(X) \log \frac{c(X)}{c'(X)} - c(X) + c'(X) \right )
\]
\end{quote}
This is most famously used in information theory to measure the information gained by moving from a prior distribution, $c'$, to a posterior, $c$. For arguments in its favour, see \citep{paris1990ime, paris1997dm, levinstein2012lp}.

Let's see the effect of these on Amira's and Benito's credences. First, let's use $\SED$. $\SED$ is symmetric --- that is, $\SED(c, c') = \SED(c', c)$, for all $c, c'$.\footnote{Indeed, $\SED$ is the only symmetric Bregman divergence.} Therefore, both fixing methods agree --- that is, $\Fix_{\SED_1} = \Fix_{\SED_2}$. GKL isn't symmetric. However, its fixing methods nonetheless always agree for credences defined on a partition --- that is, as we will see below, we also have $\Fix_{\GKL_1} = \Fix_{\GKL_2}$.
\begin{center}
\begin{tabular}{r|cc}
& $X$ & $\overline{X}$ \\
\hline 
Amira (original) & 0.5 & 0.1\\
Benito (original) & 0.2 & 0.6 \\
\hline
Amira (SED-fixed) & 0.7 & 0.3 \\
Benito (SED-fixed) & 0.3 & 0.7 \\
\hline 
Amira (GKL-fixed) & 0.83 & 0.17 \\
Benito (GKL-fixed) & 0.25 & 0.75 \\
\end{tabular}
\end{center}
%Note that our choice of divergence makes a significant difference: originally, Amira is more confident of $X$ than Benito is; using $\SED$, she remains the more confident of the two after they have both been fixed; using $\GKL$, she does not.
In general:
\begin{proposition}\label{fix-sed-gkl} For all $c$ in $C_\F$ and $X_j$ in $\F$,
\begin{enumerate}
\item[\emph{(i)}] $\Fix_{\SED_1}(c)(X_j) = \Fix_{\SED_2}(c)(X_j) = c(X_j) + \frac{1 - \sum^m_{i=1} c(X_i)}{m}$
\item[\emph{(ii)}] $\Fix_{\GKL_1}(c)(X_j) = \Fix_{\GKL_2}(c)(X_j) = \frac{c(X_j)}{\sum^m_{i=1} c(X_i)}$
\end{enumerate}
\end{proposition}
In other words, when we use $\SED$ to fix an incoherent credence function $c$ over a partition $X_1, \ldots, X_m$, we add the same quantity to each credence. That is, there is $K$ such that $\Fix_{\SED}(c)(X_j) = c(X_j) + K$, for all $1 \leq j \leq m$. Thus, the difference between a fixed credence and the original credence is always the same --- it is $K$. In order to ensure that the result is coherent, this quantity must be $K = \frac{1 - \sum^m_{i=1} c(X_i)}{m}$.  On the other hand, when we use $\GKL$ to fix $c$, we multiply each credence by the same quantity. That is, there is $K$ such that $\Fix_\GKL(c)(X_j) = K \cdot c(X_j)$, for all $1 \leq j \leq m$. Thus, the ratio between a fixed credence and the original credence is always the same --- it is $K$. In order to ensure that the result is coherent in this case, this quantity must be $K = \frac{1}{\sum^m_{i=1} c(X_i)}$.

\begin{figure}

\center

\begin{tikzpicture}[scale=2.6]
%\filldraw [green!5] (3, 0) -- (5, 0) -- (5, 5) -- (3, 5);
%\filldraw [green!5] (2, 0) -- (0, 0) -- (0, 5) -- (2, 5);
%\draw (5, 3.354) arc (90:180:3.354);
%\draw (0, 0.97) arc (270:360:4.03);
%\filldraw [blue!25] (3, 2.5) .. controls (2.7, 2.3) .. (2.5, 2);
%\filldraw [blue!25] (3, 2.5) .. controls (2.8, 2.2) .. (2.5, 2);
%\draw[densely dotted] (0, 1.5) -- (5, 1.5);
%\draw[densely dotted] (0, 3.5) -- (5, 3.5);
%\draw[densely dotted] (1.5, 0) -- (1.5, 5);
%\draw[densely dotted] (3.5, 0) -- (3.5, 5);
\filldraw  (5, 0) circle (1pt) node[anchor=north west] {$v_X$};
\filldraw  (2.5, 0) circle (0pt) node[anchor=north] {$X$};

\filldraw  (0, 2.5) circle (0pt) node[anchor=east] {$\overline{X}$};
\filldraw  (0, 5) circle (1pt) node[anchor=south east] {$v_{\overline{X}}$};

%\draw (1.3, 4.2) circle (0pt) node[anchor=west] {$\B(c, w_1)$};
%\draw (3.8, 1.8) circle (0pt) node[anchor=west] {$\B(c, w_2)$};
  \draw (-0.5,0) -- (5.5,0);
  \draw (0,-0.5) -- (0,5.5);
    \draw (-0.5,5) -- (5.5,5);
  \draw (5,-0.5) -- (5,5.5);
    
\draw [dotted] (0, 5) -- (5, 0);
%\draw[dashed] (3, 0) node[anchor=north] {0.6} -- (3, 5);

\filldraw [black] (2.5, 0.5) circle (1pt) node[anchor=north west] {$c_A$};
\filldraw [black] (1, 3) circle (1pt) node[anchor=south east] {$c_B$};
%\filldraw [black] (3.5, 1.5) circle (1pt) node[anchor=south west] {$\Fix_\SED(c_A)$};
%\filldraw [black] (1.5, 3.5) circle (1pt) node[anchor=south west] {$\Fix_\SED(c_B)$};
%\filldraw [black] (2.3, 2.7) circle (1pt) node[anchor=south west] {\begin{tabular}{l}$\LP^{\{0.4, 0.6\}}(\Fix_\SED(c_A), \Fix_\SED(c_B)) = $ \\ $\Fix_\SED(\LP^{\{0.4, 0.6\}}(c_A, c_B))$ \end{tabular} };
\filldraw [black] (2.48, 2.52) circle (1pt) node[anchor=west] {\begin{tabular}{l} \ \\ \ \ $\GP^{\{0.4, 0.6\}}(\Fix_{\GKL}(c_A), \Fix_{\GKL}(c_B))$ \\ \ \ \ \ \ \ \ \ $= \Fix_\GKL(\GP_-^{\{0.4, 0.6\}}(c_A, c_B))$ \end{tabular}};
\filldraw [black] (4.1667, 0.833) circle (1pt) node[anchor=south west] {$\Fix_{\GKL}(c_A)$};
\filldraw [black] (1.25, 3.75) circle (1pt) node[anchor=south west] {$\Fix_{\GKL}(c_B)$};
%\filldraw [black] (1.6, 2) circle (1pt) node[anchor=south east] {$\LP^{\{0.4, 0.6\}}(c_A, c_B)$};

\draw [->-=.5] plot [smooth, tension=0.1] coordinates { (1, 3) (1.09, 2.508) (1.2011, 2.096) (1.316, 1.753) (1.443, 1.465)};
% (1.581, 1.225) (1.733, 1.024) (1.899, 0.856) (2.08, 0.715) (2.281, 0.598) (2.5, 0.5)};

%\draw [-<-=.5] plot [smooth, tension=0.1] coordinates { (1.443, 1.465) (1.581, 1.225) (1.733, 1.024) (1.899, 0.856) (2.08, 0.715) (2.281, 0.598) (2.5, 0.5)};

\draw [->-=.5] plot [smooth, tension=0.1] coordinates { (2.5, 0.5) (2.281, 0.598) (2.08, 0.715) (1.899, 0.856) (1.733, 1.024) (1.581, 1.225) (1.443, 1.465) };

\filldraw [black] (1.443, 1.465) circle (1pt) node[anchor=east] {$\GP_-^{\{0.4, 0.6\}}(c_A, c_B)$\ \ };
%\filldraw [black] (1.316, 1.753) circle (1pt) node[anchor=north west] {$\GP_-^{\{0.3, 0.7\}}(c_A, c_B)$};
%\filldraw [black] (1.2011, 2.096) circle (1pt) node[anchor=north west] {$\GP_-^{\{0.2, 0.8\}}(c_A, c_B)$};
%\filldraw [black] (1.09, 2.508) circle (1pt) node[anchor=north west] {$\GP_-^{\{0.1, 0.9\}}(c_A, c_B)$};
%\filldraw [black] (1.581, 1.225) circle (1pt) node[anchor=north west] {$\GP_-^{\{0.5, 0.5\}}(c_A, c_B)$};
%\filldraw [black] (1.733, 1.024) circle (1pt) node[anchor=north west] {$\GP_-^{\{0.6, 0.4\}}(c_A, c_B)$};

%\draw[gray] (2.5, 0.5) -- (1, 3);
%\draw[densely dotted] (0, 0) -- (1, 3);
%\draw[densely dotted] (0, 0) -- (1.443, 1.465);
%\draw[densely dotted] (0, 0) -- (2.5, 0.5);
%\draw[gray!80] (2.5, 0.5) -- (1, 3);
%\draw[gray!40] (-0.5, 1.5) -- (3.5, 5.5);
%\draw[gray!40] (-0.5, -0.1) -- (5.1, 5.5);
%\draw[gray!40] (1.5, -0.5) -- (5.5, 3.5);
\draw[gray!40] (-0.1667, -0.5) -- (1.833, 5.5);
\draw[gray!40] (-0.492, -0.5) -- (5.412, 5.5);
\draw[gray!40] (-0.5, -0.1) -- (5.5, 1.1);

\draw [->-=0.5] (1.25, 3.75) -- (2.48, 2.52);
\draw [->-=0.5] (4.1667, 0.833) -- (2.48, 2.52);
\draw [->-=0.5] (1, 3) -- (1.25, 3.75);
\draw [->-=0.5] (2.5, 0.5) -- (4.1667, 0.833);
\draw [->-=0.5] (1.443, 1.465) -- (2.48, 2.52);

%\draw[black] (1,3) -- (1.5,3.5);
    % \draw[black] (1,3) -- (1.25,3.75);
   %  \draw[black] (2.5,0.5) -- (3.5,1.5);
     %\draw[black] (2.5,0.5) -- (4.1667, 0.833);
     %\draw[black] (2.3, 2.7) -- (1.6, 2);
   %  \draw[black] (2.48, 2.52) -- (1.443, 1.465);

%\draw[densely dotted] (3, 3) -- (5, 0);

%\filldraw [black] (2.5, 2.5) circle (1pt) node[anchor=north] {$c^*$};
%\filldraw [black] (3, 2) circle (1pt) node[anchor=south west] {$c'$};
\end{tikzpicture}

%\begin{tikzpicture}[scale=1.3]
%\filldraw  (5, 0) circle (1pt) node[anchor=north west] {$v_{w_2}$};

%\filldraw (0, 5) circle (1pt) node[anchor=south east] {$v_{w_1}$};
%\draw (1.3, 4.2) circle (0pt) node[anchor=west] {$\B(c, w_1)$};
%\draw (3.8, 1.8) circle (0pt) node[anchor=west] {$\B(c, w_2)$};
%   \draw (-0.5,0) -- (5.5,0) ;%node[anchor=north] {$A$};
%  \draw (0,-0.5) -- (0,5.5);
%  \draw (-0.5,5) -- (5.5,5);
%  \draw (5,-0.5) -- (5,5.5);
%\draw[thick]  (0, 5) -- (5, 0);
%\draw[dotted] (3, 0) node[anchor=north] {0.6} -- (3, 5);
%\filldraw [black] (3, 3) circle (1pt) node[anchor=west] {$c$};
%\draw[densely dotted] [red] (3, 3) -- (0, 5);
%\draw[densely dotted] [red] (3, 3) -- (5, 0);
%\filldraw [black!25] (3, 3) .. controls (2.4505, 2.5495) .. (2, 2);
%\filldraw [black!25] (3, 3) .. controls (2.5495, 2.4505) .. (2, 2);
%\filldraw [black] (2.5, 2.5) circle (1pt) node[anchor=north] {$c^*$};
%\filldraw [black] (3, 2) circle (1pt) node[anchor=south west] {$c'$};
%\end{tikzpicture}
\caption{\label{amira-benito-gkl} Here, we see that $\Fix_\GKL(c_A)$ is the projection from the origin through $c_A$ onto the set of coherent credence functions; and similarly for $\Fix_\GKL(c_B)$ and $\Fix_\GKL(\GP_-^{\{0.4, 0.6\}}(c_A, c_B))$. The curved line from $c_A$ to $c_B$ represents the set of geometric pools of $c_A$ and $c_B$ generated by different weightings. Again, the arrows indicate that $\GP = \Fix_\GKL \circ \GP_- = \GP \circ \Fix_\GKL$.}
\end{figure}
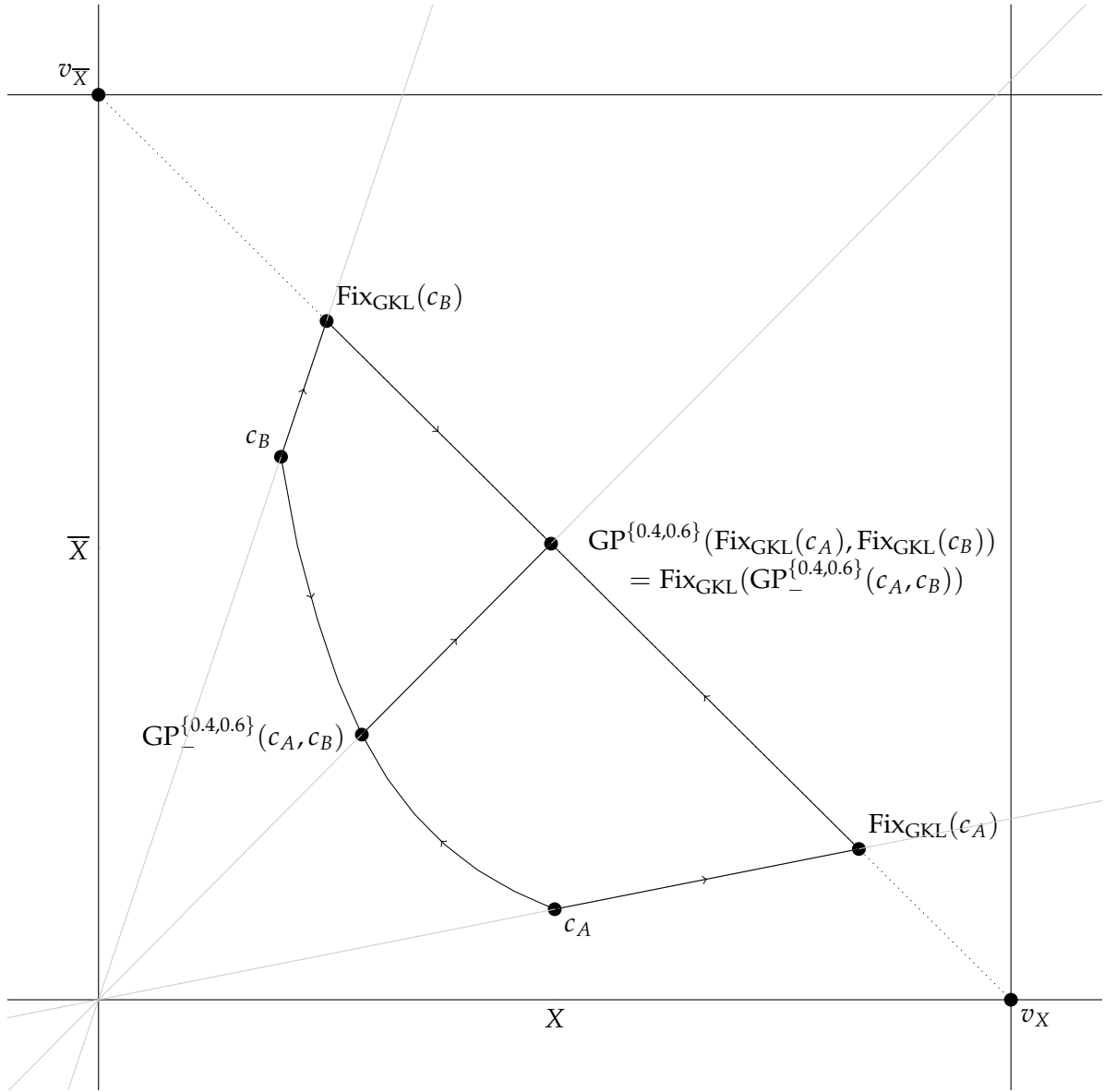

There is also a geometric way to understand the relationship between fixing using $\SED$ and fixing using $\GKL$. Roughly: $\Fix_\SED(c)$ is the orthogonal projection of $c$ onto the set of coherent credence functions, while $\Fix_\GKL(c)$ is the result of projecting from the origin through $c$ onto the set of coherent credence functions. This is illustrated in Figures \ref{amira-benito-sed} and \ref{amira-benito-gkl}. One consequence is this: if $c(X) + c(\overline{X}) < 1$, then fixing using $\SED$ is more conservative than fixing by $\GKL$, in the sense that the resulting credence function is less opinionated --- it has a lower maximum credence. But if $c(X) + c(\overline{X}) > 1$, then fixing using $\GKL$ is more conservative.

\section{Aggregate-then-fix vs fix-then-aggregate}\label{agg-fix-fix-agg}

Using the formulae in Proposition \ref{fix-sed-gkl}, we can explore what differences, if any, there are between fixing incoherent agents and then aggregating them, on the one hand, and aggregating incoherent agents and then fixing the aggregate, on the other. Suppose $c_1$, \ldots, $c_n$ are the credence functions of a group of agents, all defined on the same partition $X_1, \ldots, X_m$. Some may be incoherent, and we wish to aggregate them. Thus, we might first fix each credence function, and then aggregate the resulting coherent credence functions; or we might aggregate the original credence functions, and then fix the resulting aggregate. When we aggregate, we have two methods at our disposal --- linear pooling (LP) and geometric pooling (GP); and when we fix, we have two methods at our disposal --- one based on squared Euclidean distance ($\SED$) and the other based on generalized Kullback-Leibler divergence ($\GKL$). Our next result tells us how these different options interact. To state it, we borrow a little notation from the theory of  function composition. For instance, we write $\LP^{\{\alpha\}} \circ \Fix_\SED$ to denote the function that takes a collection of agents' credence functions $c_1$, \ldots, $c_n$ and returns $\LP^{\{\alpha\}}(\Fix_\SED(c_1), \ldots,  \Fix_\SED(c_n))$. So $\LP^{\{\alpha\}} \circ \Fix_\SED$ might be read: \emph{$\LP^{\{\alpha\}}$ following $\Fix_\SED$}, or \emph{$\LP^{\{\alpha\}}$ acting on the results of $\Fix_\SED$}. Similarly, $\Fix_\SED \circ \LP^{\{\alpha\}}$ denotes the function that takes $c_1$, \ldots, $c_n$ and returns $\Fix_\SED(\LP^{\{\alpha\}}(c_1, \ldots, c_n))$. And we say that two functions are equal if they agree on all arguments, and unequal if they disagree on some.
\begin{proposition}\label{commutes}\ 
\begin{enumerate}
\item[\emph{(i)}] $\LP^{\{\alpha\}} \circ \Fix_\SED = \Fix_\SED \circ \LP^{\{\alpha\}}$.

That is, linear pooling commutes with $\SED$-fixing.

That is, for all $c_1$, \ldots $c_n$,
\[
\LP^{\{\alpha\}}(\Fix_\SED(c_1), \ldots,  \Fix_\SED(c_n)) = \Fix_\SED(\LP^{\{\alpha\}}(c_1, \ldots, c_n))
\]
And the resulting credence function assigns the following credence to $X_j$:
\[
\sum^n_{k=1} \alpha_k c_k(X_j) + \frac{1 - \sum^m_{i=1} \sum^n_{k=1} \alpha_k c_k(X_i)}{m}
\]
\item[\emph{(ii)}] $\LP^{\{\alpha\}} \circ \Fix_\GKL \neq \Fix_\GKL \circ \LP^{\{\alpha\}}$.

That is, linear pooling does not commute with $\GKL$-fixing.

That is, for some $c_1$, \ldots, $c_n$,
\[
\LP^{\{\alpha\}}(\Fix_\GKL(c_1), \ldots, \Fix_\GKL(c_n)) \neq \Fix_\GKL(\LP^{\{\alpha\}}(c_1, \ldots, c_n))
\]
\item[\emph{(iii)}] $\GP^{\{\alpha\}} \circ \Fix_\GKL = \Fix_\GKL \circ \GP^{\{\alpha\}}$.

That is, geometric pooling commutes with $\GKL$-fixing.

That is, for all $c_1$, \ldots, $c_n$,
\[
\GP^{\{\alpha\}}(\Fix_\GKL(c_1), \ldots, \Fix_\GKL(c_n)) = \Fix_\GKL(\GP^{\{\alpha\}}(c_1, \ldots, c_n))
\]
And the resulting credence function assigns the following credence to $X_j$:
\[
\frac{\prod^n_{k=1} c_k(X_j)^{\alpha_k}}{\sum^m_{i=1} \prod^n_{k=1} c_k(X_i)^{\alpha_k}}
\]
\item[\emph{(iv)}] $\GP^{\{\alpha\}} \circ \Fix_\SED \neq \Fix_\SED \circ \GP^{\{\alpha\}}$.

That is, geometric pooling does not commute with $\SED$-fixing.

That is, for some $c_1$, \ldots, $c_n$,
\[
\GP^{\{\alpha\}}(\Fix_\SED(c_1), \ldots, \Fix_\SED(c_n)) \neq \Fix_\SED(\GP^{\{\alpha\}}(c_1, \ldots, c_n))
\]
\end{enumerate}
\end{proposition}
With this result, we start to see the main theme of this paper emerging: $\SED$ naturally accompanies linear pooling, while $\GKL$ naturally accompanies geometric pooling. In section \ref{main-results}, we'll present further results that support that conclusion, as well as some that complicate it a little.

\section{Aggregating by minimizing distance}\label{agg-min-dist}

In the previous section, we introduced the notion of a divergence and we put it to use fixing incoherent credence functions: given an incoherent credence function $c$, we fix it by taking the coherent credence function that minimizes divergence \emph{to} or \emph{from} $c$. But divergences can also be used to aggregate credence functions.\footnote{When the agents are represented as having categorical doxastic states, such as full beliefs or commitments, this method was studied first in the computer science literature on \emph{belief merging} \citep{konieczny1999mic, konieczny2006lba}. It was studied first in the judgment aggregation literature by Gabriella \citet{pigozzi2006bm}.} The idea is this: given a divergence and a collection of credence functions, take the aggregate to be the credence function that minimizes the weighted arithmetic average of the divergences \emph{to} or \emph{from} those credence functions. Thus:
\begin{quote} \textbf{$\D$-aggregation } Let $\{\alpha\}$ be a set of weights. Then
\[
\Agg^{\{\alpha\}}_{\D_1}(c_1, \ldots, c_n) = \underset{c' \in C_\F}{\mathrm{arg\, min}\ } \sum^n_{k=1} \alpha_k \D(c', c_k)
\]
and
\[
\Agg^{\{\alpha\}}_{\D_2}(c_1, \ldots, c_n) = \underset{c' \in C_\F}{\mathrm{arg\, min}\ } \sum^n_{k=1}\alpha_k \D(c_k, c')
\]
\end{quote}
Let's see what these give when applied to the two divergences we introduced above, namely, $\SED$ and $\GKL$.
\begin{proposition}\label{aggregate} Let $\{\alpha\}$ be a set of weights. Then, for each $X_j$ in $\F$,
\begin{enumerate}
\item[\emph{(i)}] 
\[
\Agg^{\{\alpha\}}_{\SED}(c_1, \ldots, c_n)(X_j) = \sum^n_{k=1} \alpha_k c_k(X_j) = \LP^{\{\alpha\}}(c_1, \ldots, c_n)
\]
\item[\emph{(ii)}] 
\[
\Agg^{\{\alpha\}}_{\GKL_1}(c_1, \ldots, c_n)(X_j) = \prod^n_{k=1} c_k(X_j)^{\alpha_k} =  \GP_-^{\{\alpha\}}(c_1, \ldots, c_n)
\]
\item[\emph{(ii)}] 
\[
\Agg^{\{\alpha\}}_{\GKL_2}(c_1, \ldots, c_n)(X_j) = \sum^n_{k=1} \alpha_k c_k(X_j) =  \LP^{\{\alpha\}}(c_1, \ldots, c_n)
\]
\end{enumerate}
\end{proposition}
Thus, $\Agg_\SED$ and $\Agg_{\GKL_2}$ are just linear pooling --- they assign to each $X_j$ the (unnormalized) weighted arithmetic average of the credences assigned to $X_j$ by the agents. On the other hand, $\Agg_{\GKL_1}$ is just geometric pooling without the normalization procedure --- it assigns to each $X_j$ the (unnormalized) weighted geometic average of the credences assigned to $X_j$ by the agents. I call this aggregation procedure $\GP_-$. Given a set of coherent credence functions, $\GP$ returns a coherent credence function, but $\GP_-$ typically won't. However, if we aggregate using $\GP_-$ and then fix using $\GKL$, then we obtain $\GP$:
\begin{proposition} Let $\{\alpha\}$ be a set of weights. Then
\[
\GP^{\{\alpha\}} = \Fix_\GKL \circ \GP_-^{\{\alpha\}}
\]
\end{proposition}

%Finally, let's see how these interact with the corresponding ways of fixing incoherent credence functions:
%\begin{proposition}\label{d-agg} Let $\{\alpha\}$ be a set of weights. Then
%\begin{enumerate}
%\item[\emph{(i)}] $\Agg_\SED^{\{\alpha\}} \circ \Fix_\SED = \Fix_\SED \circ \Agg_\SED^{\{\alpha\}} = \LP^{\{\alpha\}} \circ \Fix_\SED = \Fix_\SED \circ \LP^{\{\alpha\}}$.

%That is, $\Agg_\SED$ commutes with $\SED$-fixing, and their combination is the same as linear pooling-then-fixing or fixing-then-linear-pooling.
%\item[\emph{(ii)}] $\Agg_{\GKL_1}^{\{\alpha\}} \circ \Fix_{\GKL} = \Fix_\GKL \circ \Agg_{\GKL_1}^{\{\alpha\}} = \GP^{\{\alpha\}}$.

%That is, $\Agg_{\GKL_1}$ commutes with $\GKL$-fixing, and their combination is the same as geometric  pooling.
%\item[\emph{(iii)}] $\Agg_{\GKL_2}^{\{\alpha\}} \circ \Fix_{\GKL} \neq \Fix_\GKL \circ \Agg_{\GKL_2}^{\{\alpha\}}$.

%That is, $\Agg_{\GKL_2}$ does not commute with $\GKL$-fixing.
%\end{enumerate}
%\end{proposition}

\section{Aggregate and fix together}\label{agg-fix-together}

In this section, we meet our final procedure for producing a single coherent credence function from a collection of possibly incoherent ones. This procedure fixes and aggregates together: that is, it is a one-step process, unlike the two-step processes we have considered so far. It generalises a technique suggested by \citet{osherson2006ade} and explored further by \citet{predd2008apf}.\footnote{\citet{osherson2006ade} and \citet{predd2008apf} consider only what we call $\WCAP^{\{\frac{1}{n}\}}_{\SED_1}$ below. They do not consider the different Bregman divergences $\D$; they do not consider the two directions; and they do not consider the possibility of weighting the distances differently. This is quite understandable --- their interest lies mainly in the feasibility of the procedure from a computational point of view. We will not address this issue here.} Again, it appeals to a divergence $\D$; and thus again, there are two versions, depending on whether we measure distance \emph{from} coherence or distance \emph{to} coherence. If we measure distance from coherence, the \emph{weighted coherent approximation principle} tells us to pick the coherent credence function such that the weighted arithmetic average of the divergences from that coherent credence function to the agents is minimized. And if we measure distance to coherence, it picks the coherent credence function that minimizes the weighted arithmetic average of the divergences from the agents to the credence function. Thus, it poses a minimization problem similar to that posed by $\D$-aggregation, but in this case, we wish to find the credence function \emph{amongst the coherent ones} that does the minimizing; in the case of $\D$-aggregation, we wish to find the credence function \emph{amongst all the credence ones} that does the minimizing.
\begin{quote}
\textbf{Weighted Coherent Approximation Principle } Let $\{\alpha\}$ be a set of weights. Then

\[
\WCAP^{\{\alpha\}}_{\D_1}(c_1, \ldots, c_n) = \underset{c' \in P_\F}{\mathrm{arg\, min}\ } \sum^n_{k=1}\alpha_k \D(c', c_k)
\]
and
\[
\WCAP^{\{\alpha\}}_{\D_2}(c_1, \ldots, c_n) = \underset{c' \in P_\F}{\mathrm{arg\, min}\ } \sum^n_{k=1}\alpha_k \D(c_k, c')
\]
\end{quote}
How does this procedure compare to the fix-then-aggregate and aggregate-then-fix procedures that we considered above? Our next result gives the answer:
\begin{proposition}\label{sed-gkl} Let $\{\alpha\}$ be a set of weights. Then
\begin{enumerate}
\item[\emph{(i)}]
\[
\begin{array}{lllll}
\WCAP^{\{\alpha\}}_\SED & = & \LP^{\{\alpha\}} \circ \Fix_\SED & = & \Fix_\SED \circ \LP^{\{\alpha\}} \\
& = & \Agg_\SED^{\{\alpha\}} \circ \Fix_\SED & = & \Fix_\SED \circ \Agg_\SED^{\{\alpha\}}
\end{array}
\]
\item[\emph{(ii)}]
\[
\begin{array}{lllll}
\WCAP^{\{\alpha\}}_{\GKL_1} & = & \GP^{\{\alpha\}} \circ \Fix_\GKL & = & \Fix_\GKL \circ \GP^{\{\alpha\}} \\
& = & \Agg_{\GKL_1}^{\{\alpha\}} \circ \Fix_\GKL & = & \Fix_\GKL \circ \Agg_{\GKL_1}^{\{\alpha\}} \\
& = & \GP^{\{\alpha\}} 
\end{array}
\]
\item[\emph{(iii)}]
\[
\begin{array}{lllll}
\WCAP^{\{\alpha\}}_{\GKL_2} & \neq & \GP^{\{\alpha\}} \circ \Fix_\GKL & = & \Fix_\GKL \circ \GP^{\{\alpha\}}\\
&  = & \GP^{\{\alpha\}}
\end{array}
\]
In this case,
\begin{eqnarray*}
\WCAP^{\{\alpha\}}_{\GKL_2}(c_1, \ldots, c_n)(X_j) & = & \frac{\sum^n_{k=1} \alpha_k c_k(X_j)}{\sum^n_{k=1} \alpha_k c_k(X_1) + \ldots + \sum^n_{k=1} \alpha_k c_k(X_n)} \\
& = & \frac{\sum^n_{k=1} \alpha_k c_k(X_j)}{\sum^m_{i=1} \sum^n_{k=1} \alpha_k c_k(X_i)}
\end{eqnarray*}
Thus, it first takes a linear pool of the credences and then normalizes.
\item[\emph{(iv)}]
\[
\begin{array}{lllll}
\WCAP^{\{\alpha\}}_{\GKL_2} & = & \Fix_\GKL \circ \Agg_{\GKL_2}^{\{\alpha\}} & \neq & \Agg_{\GKL_2}^{\{\alpha\}} \circ \Fix_\GKL
\end{array}
\]
\end{enumerate}
\end{proposition}
(i) and (ii) confirm our picture that linear pooling naturally pairs with $\SED$, while geometric pooling pairs naturally with $\GKL$. However, (iii) and (iv) complicate this. This is a pattern we will continue to see as we progress: when we miminize distance \emph{from} coherence, the aggregation methods and divergence measures pair up neatly; when we minimize distance \emph{to} coherence, they do not.

These, then, are the various ways we will consider by which you might produce a single coherent credence function when given a collection of possibly incoherent ones: fix-then-aggregate, aggregate-then-fix, and the weighted coherent approximation principle. Each involves minimizing a divergence at some point, and so each comes in two varieties, one based on minimizing distance \emph{from} coherence, the other based on minimizing distance \emph{to} coherence.

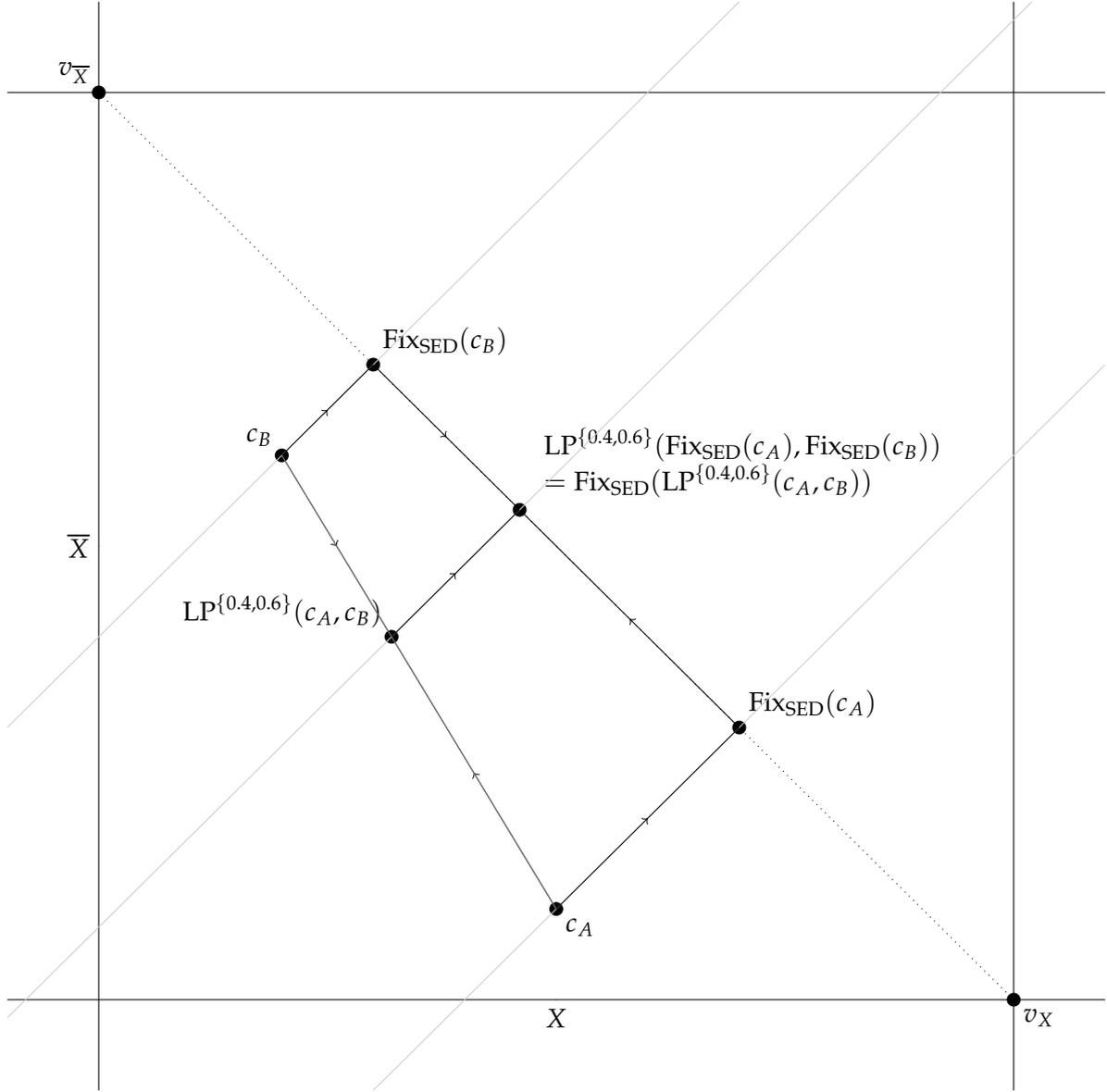
\begin{figure}

\center

\begin{tikzpicture}[scale=2.6]
%\filldraw [green!5] (3, 0) -- (5, 0) -- (5, 5) -- (3, 5);
%\filldraw [green!5] (2, 0) -- (0, 0) -- (0, 5) -- (2, 5);
%\draw (5, 3.354) arc (90:180:3.354);
%\draw (0, 0.97) arc (270:360:4.03);
%\filldraw [blue!25] (3, 2.5) .. controls (2.7, 2.3) .. (2.5, 2);
%\filldraw [blue!25] (3, 2.5) .. controls (2.8, 2.2) .. (2.5, 2);
%\draw[densely dotted] (0, 1.5) -- (5, 1.5);
%\draw[densely dotted] (0, 3.5) -- (5, 3.5);
%\draw[densely dotted] (1.5, 0) -- (1.5, 5);
%\draw[densely dotted] (3.5, 0) -- (3.5, 5);
\filldraw  (5, 0) circle (1pt) node[anchor=north west] {$v_X$};
\filldraw  (2.5, 0) circle (0pt) node[anchor=north] {$X$};

\filldraw  (0, 2.5) circle (0pt) node[anchor=east] {$\overline{X}$};
\filldraw  (0, 5) circle (1pt) node[anchor=south east] {$v_{\overline{X}}$};

%\draw (1.3, 4.2) circle (0pt) node[anchor=west] {$\B(c, w_1)$};
%\draw (3.8, 1.8) circle (0pt) node[anchor=west] {$\B(c, w_2)$};
  \draw (-0.5,0) -- (5.5,0);
  \draw (0,-0.5) -- (0,5.5);
    \draw (-0.5,5) -- (5.5,5);
  \draw (5,-0.5) -- (5,5.5);
    
\draw [dotted] (0, 5) -- (5, 0);
%\draw[dashed] (3, 0) node[anchor=north] {0.6} -- (3, 5);

\draw [->-=0.5] (3.5, 1.5) -- (2.3, 2.7);
\draw [->-=0.5] (1.5, 3.5) -- (2.3, 2.7);
\draw [->-=0.5] (1, 3) -- (1.6, 2);
\draw [->-=0.5] (2.5, 0.5) -- (1.6, 2);

\filldraw [black] (2.5, 0.5) circle (1pt) node[anchor=north west] {$c_A$};
\filldraw [black] (1, 3) circle (1pt) node[anchor=south east] {$c_B$};
\filldraw [black] (3.5, 1.5) circle (1pt) node[anchor=south west] {$\Fix_\SED(c_A)$};
\filldraw [black] (1.5, 3.5) circle (1pt) node[anchor=south west] {$\Fix_\SED(c_B)$};
\filldraw [black] (2.3, 2.7) circle (1pt) node[anchor=south west] {\begin{tabular}{l}$\LP^{\{0.4, 0.6\}}(\Fix_\SED(c_A), \Fix_\SED(c_B))$ \\ $= \Fix_\SED(\LP^{\{0.4, 0.6\}}(c_A, c_B))$ \end{tabular} };
%\filldraw [black] (2.48, 2.52) circle (1pt) node[anchor=west] {\begin{tabular}{l} \ \\ \ \ \ $\GP^{\{0.4, 0.6\}}(\Fix_{\GKL}(c_A), \Fix_{\GKL}(c_B)) = $ \\ \ \ \ \ \ \ \ \ $\Fix_\GKL(\GP_-^{\{0.4, 0.6\}}(c_A, c_B))$ \end{tabular}};
%\filldraw [black] (4.1667, 0.833) circle (1pt) node[anchor=south west] {$\Fix_{\GKL}(c_A)$};
%\filldraw [black] (1.25, 3.75) circle (1pt) node[anchor=south west] {$\Fix_{\GKL}(c_B)$};
\filldraw [black] (1.6, 2) circle (1pt) node[anchor=south east] {$\LP^{\{0.4, 0.6\}}(c_A, c_B)$};

%\draw [gray] plot [smooth, tension=0.1] coordinates { (1, 3) (1.09, 2.508) (1.2011, 2.096) (1.316, 1.753) (1.443, 1.465) (1.581, 1.225) (1.733, 1.024) (1.899, 0.856) (2.08, 0.715) (2.281, 0.598) (2.5, 0.5)};

%\filldraw [black] (1.443, 1.465) circle (1pt) node[anchor=east] {$\GP_-^{\{0.4, 0.6\}}(c_A, c_B)$\ \ };
%\filldraw [black] (1.316, 1.753) circle (1pt) node[anchor=north west] {$\GP_-^{\{0.3, 0.7\}}(c_A, c_B)$};
%\filldraw [black] (1.2011, 2.096) circle (1pt) node[anchor=north west] {$\GP_-^{\{0.2, 0.8\}}(c_A, c_B)$};
%\filldraw [black] (1.09, 2.508) circle (1pt) node[anchor=north west] {$\GP_-^{\{0.1, 0.9\}}(c_A, c_B)$};
%\filldraw [black] (1.581, 1.225) circle (1pt) node[anchor=north west] {$\GP_-^{\{0.5, 0.5\}}(c_A, c_B)$};
%\filldraw [black] (1.733, 1.024) circle (1pt) node[anchor=north west] {$\GP_-^{\{0.6, 0.4\}}(c_A, c_B)$};

\draw[gray] (2.5, 0.5) -- (1, 3);
%\draw[densely dotted] (0, 0) -- (1, 3);
%\draw[densely dotted] (0, 0) -- (1.443, 1.465);
%\draw[densely dotted] (0, 0) -- (2.5, 0.5);
%\draw[gray!80] (2.5, 0.5) -- (1, 3);
\draw[gray!40] (-0.5, 1.5) -- (3.5, 5.5);
\draw[gray!40] (-0.5, -0.1) -- (5.1, 5.5);
\draw[gray!40] (1.5, -0.5) -- (5.5, 3.5);
%\draw[gray!40] (-0.1667, -0.5) -- (1.833, 5.5);
%\draw[gray!40] (-0.492, -0.5) -- (5.412, 5.5);
%\draw[gray!40] (-0.5, -0.1) -- (5.5, 1.1);

\draw [->-=.5] (1,3) -- (1.5,3.5);
    % \draw[black] (1,3) -- (1.25,3.75);
     \draw [->-=.5]  (2.5,0.5) -- (3.5,1.5);
    % \draw[black] (2.5,0.5) -- (4.1667, 0.833);
     \draw [->-=.5]  (1.6, 2) -- (2.3, 2.7) ;
    % \draw[black] (2.48, 2.52) -- (1.443, 1.465);

%\draw[densely dotted] (3, 3) -- (5, 0);

%\filldraw [black] (2.5, 2.5) circle (1pt) node[anchor=north] {$c^*$};
%\filldraw [black] (3, 2) circle (1pt) node[anchor=south west] {$c'$};
\end{tikzpicture}

%\begin{tikzpicture}[scale=1.3]
%\filldraw  (5, 0) circle (1pt) node[anchor=north west] {$v_{w_2}$};

%\filldraw (0, 5) circle (1pt) node[anchor=south east] {$v_{w_1}$};
%\draw (1.3, 4.2) circle (0pt) node[anchor=west] {$\B(c, w_1)$};
%\draw (3.8, 1.8) circle (0pt) node[anchor=west] {$\B(c, w_2)$};
%   \draw (-0.5,0) -- (5.5,0) ;%node[anchor=north] {$A$};
%  \draw (0,-0.5) -- (0,5.5);
%  \draw (-0.5,5) -- (5.5,5);
%  \draw (5,-0.5) -- (5,5.5);
%\draw[thick]  (0, 5) -- (5, 0);
%\draw[dotted] (3, 0) node[anchor=north] {0.6} -- (3, 5);
%\filldraw [black] (3, 3) circle (1pt) node[anchor=west] {$c$};
%\draw[densely dotted] [red] (3, 3) -- (0, 5);
%\draw[densely dotted] [red] (3, 3) -- (5, 0);
%\filldraw [black!25] (3, 3) .. controls (2.4505, 2.5495) .. (2, 2);
%\filldraw [black!25] (3, 3) .. controls (2.5495, 2.4505) .. (2, 2);
%\filldraw [black] (2.5, 2.5) circle (1pt) node[anchor=north] {$c^*$};
%\filldraw [black] (3, 2) circle (1pt) node[anchor=south west] {$c'$};
%\end{tikzpicture}
\caption{\label{amira-benito-sed} \emph{Linear pooling and $\SED$-fixing applied to Amira's and Benito's credences.} If $\F = \{X, \overline{X}\}$, we can represent the set of all credence functions defined on $X$ and $\overline{X}$ as the points in the unit square: we represent $c : \{X, \overline{X}\} \rightarrow [0, 1]$ as the point $(c(X), c(\overline{X}))$, so that the $x$-coordinate gives the credence in $X$, while the $y$-coordinate gives the credence in $\overline{X}$. In this way, we represent Amira's credence function as $c_A$ and Benito's as $c_B$ in the diagram above. And $\mathcal{P}_\F$, the set of coherent credence functions, is represented by the thick diagonal line joining the omniscient credence functions $v_X$ and $v_{\overline{X}}$. As we can see, $\Fix_\SED(c_A)$ is the orthogonal projection of $c_A$ onto this set of coherent credence functions; and similarly for $\Fix_\SED(c_B)$ and $\Fix_\SED(\LP^{\{0.4, 0.6\}}(c_A, c_B))$. The straight line from $c_A$ to $c_B$ represents the set of linear pools of $c_A$ and $c_B$ generated by different weightings. The arrows indicated that you can reach the same point --- $\LP^{\{0.4, 0.6\}}(\Fix_\SED(c_A), \Fix_\SED(c_B)) = \Fix_\SED(\LP^{\{0.4, 0.6\}}(c_A, c_B))$ --- from either direction. That is, $\LP$ and $\Fix_\SED$ commute.}
\end{figure}

Are these the only possible ways? There is one other that might seem a natural cousin of $\WCAP$, and one that might seem a natural cousin of $\D$-aggregation, which we might combine with fixing in either of the ways considered above. In $\WCAP$, we pick the coherent credence function that minimizes the weighted \emph{arithmetic} average of the distances from (or to) the agents. The use of the weighted arithmetic average here might lead you to expect that $\WCAP$ will pair most naturally with linear pooling, which aggregates by taking the weighted arithmetic average of the agents' credences. You might expect it to interact poorly with geometric pooling, which aggregates by taking the weighted geometric average of the agents' credences (and then normalizing). But, in fact, as we saw in Theorem \ref{sed-gkl}(ii), when coupled with the divergence $\GKL$, and when we minimize distance from coherence, rather than distance to coherence, $\WCAP$ entails geometric pooling. Nonetheless, we might think that if it is natural to minimize the weighted \emph{arithmetic} average of distances from coherence, and if both linear and geometric pooling are on the table, revealing that we have no prejudice against using geometric averages to aggregate numerical values, then it is equally natural to minimize the weighted \emph{geometric} average of distances from coherence. This gives:
\begin{quote}
\textbf{Weighted Geometric Coherent Approximation Principle } Let $\{\alpha\}$ be a set of weights. Then
\[
\WGCAP^{\{\alpha\}}_{\D_1}(c_1, \ldots, c_n) = \underset{c' \in P_\F}{\mathrm{arg\, min}\ } \prod^n_{k=1}\D(c', c_k)^{\alpha_k}
\]
and
\[
\WGCAP^{\{\alpha\}}_{\D_2}(c_1, \ldots, c_n) = \underset{c' \in P_\F}{\mathrm{arg\, min}\ } \prod^n_{k=1}\D(c_k, c')^{\alpha_k}
\]
\end{quote}
However, it is easy to see that:
\begin{proposition}\label{dictator-1}
For any divergence $\D$, any set of weights $\{\alpha\}$, any $i \in \{1, 2\}$, and any coherent credence functions $c_1$, \ldots, $c_n$,
\[
c^* = \WGCAP^{\{\alpha\}}_{\D_i}(c_1, \ldots, c_n) \mbox{\ \ \ iff\ \ \ } c^* = c_1 \mbox{ or \ldots or } c_n
\]
\end{proposition}
That is, $\WGCAP$ gives a dictatorship rule when applied to coherent agents: it aggregates a group of agents by picking one of those agents and making her stand for the whole group. This rules it out immediately as a method of aggregation.

Similarly, we might define a geometric cousin to $\D$-aggregation:
\begin{quote}
\textbf{Geometric $\D$-aggregation } Let $\{\alpha\}$ be a set of weights. Then
\[
\GAgg^{\{\alpha\}}_{\D_1}(c_1, \ldots, c_n) = \underset{c' \in C_\F}{\mathrm{arg\, min}\ } \prod^n_{k=1}\D(c', c_k)^{\alpha_k}
\]
and
\[
\GAgg^{\{\alpha\}}_{\D_2}(c_1, \ldots, c_n) = \underset{c' \in C_\F}{\mathrm{arg\, min}\ } \prod^n_{k=1}\D(c_k, c')^{\alpha_k}
\]
\end{quote}
However, we obtain a similar result to before, though this time the dictatorship arises for \emph{any} set of agents, not just coherent ones.
\begin{proposition}\label{dictator-2}
For any divergence $\D$, any set of weights $\{\alpha\}$, any $i \in \{1, 2\}$, and any credence functions $c_1$, \ldots, $c_n$,
\[
c^* = \GAgg^{\{\alpha\}}_{\D_i}(c_1, \ldots, c_n) \mbox{\ \ \ iff\ \ \ } c^* = c_1 \mbox{ or \ldots or } c_n
\]
\end{proposition}
Thus, in what follows, we will consider only fix-then-aggregate, aggregate-then-fix, and $\WCAP$.

\section{Bregman divergences}\label{bregman}

In the previous section, we stated our definition of \emph{fixing} and our definition of the \emph{weighted coherent approximation principle} in terms of a divergence $\D$. We then identified two such divergences, $\SED$ and $\GKL$, and we explored how those ways of making incoherent credences coherent related to ways of combining different credence functions to give a single one. This leaves us with two further questions: Which other divergences might we use when we are fixing incoherent credences? And how do the resulting ways of fixing relate to our aggregation principles? In this section, we introduce a large family of divergences known as the \emph{additive Bregman divergences} \citep{bregman1967rm}. $\SED$ and $\GKL$ are both additive Bregman divergences, and indeed Bregman introduced the notion as a generalisation of $\SED$. They are widely used in statistics to measure how far one probability distribution lies from another \citep{csiszar1991lsme, banerjee2005opt, gneiting2007spsr, csiszar2008acim, predd2009pc}; they are used in social choice theory to measure how far one distribution of wealth lies from another \citep{dagostino2009wss, magdalou2011id}; and they are used in the epistemology of credences to define measures of the inaccuracy of credence functions \citep{pettigrew2016alc}. Below, I will offer some reasons why we should use them in our procedures for fixing incoherent credences. But first let's define them.

Each additive Bregman divergence $\D : C_\F \times C_\F \rightarrow [0, \infty]$ is generated by a function $\varphi : [0, 1] \rightarrow \mathbb{R}$, which is required to be (i) strictly convex on $[0, 1]$ and (ii) twice differentiable on (0, 1) with a continuous second derivative. We begin by using $\varphi$ to define the divergence from $x$ to $y$, where $0 \leq x, y \leq 1$. We first draw the tangent to $\varphi$ at $y$. Then we take the divergence from $x$ to $y$ to be the difference between the value of $\varphi$ at $x$ --- that is, $\varphi(x)$ --- and the value of that tangent at $x$ --- that is, $\varphi(y) + \varphi'(y)(x - y)$. Thus, the divergence from $x$ to $y$ is $\varphi(x) - \varphi(y) - \varphi'(y)(x-y)$. We then take the divergence from one credence function $c$ to another $c'$ to be the sum of the divergences from each credence assigned by $c$ to the corresponding credence assigned by $c'$. Thus:
\begin{definition}
Suppose $\varphi: [0, 1] \rightarrow \mathbb{R}$ is a strictly convex function that is twice differentiable on $(0, 1)$ with a continuous second derivative. And suppose $\D : C_\F \times C_\F \rightarrow [0, \infty]$. Then \emph{$\D$ is the additive Bregman divergence generated by $\varphi$} if, for any $c, c'$ in $C_\F$,
\[
\D(c, c') = \sum^m_{i=1} \varphi(c(X_i)) - \varphi(c'(X_i)) - \varphi'(c'(X_i))(c(X_i) - c'(X_i))
\]
\end{definition}
And we can show:
\begin{proposition}\label{bregman-sed-gkl}\ 
\begin{enumerate}
\item[\emph{(i)}] $\SED$ is the additive Bregman divergence generated by $\varphi(x) = x^2$.
\item[\emph{(ii)}] $\GKL$ is the additive Bregman divergence generated by $\varphi(x) = x \log x - x$.
\end{enumerate}
\end{proposition}

Why do we restrict our attention to additive Bregman divergences when we are considering which divergences to use to fix incoherent credences? Here's one answer.\footnote{See \citep{debona2016gi} for a similar line of argument.} Just as beliefs can be true or false, credences can be more or less accurate. A credence in a true proposition is more accurate the higher it is, while a credence in a false proposition is more accurate the lower it is. Now, just as some philosophers think that beliefs are more valuable if they are true than if they are false \citep{goldman2002ptk}, so some philosophers think that credences are more valuable the more accurate they are \citep{joyce1998nvp, pettigrew2016alc}. This approach is sometimes called \emph{accuracy-first epistemology}. These philosophers then provide mathematically precise ways to measure the \emph{in}accuracy of credence functions. They say that a credence function $c$ is more inaccurate at a possible world $w$ the further $c$ lies from the omniscient credence function $v_w$ at $w$, where $v_w$ assigns maximal credence (i.e. 1) to all truths at $w$ and minimal credence (i.e. 0) to all falsehoods at $w$. So, in order to measure the inaccuracy of $c$ at $w$ we need a measure of how far one credence function lies from another, just as we do when we want to fix incoherent credence functions. But which divergences are legitimate measures for this purpose? Elsewhere, I have argued that it is only the additive Bregman divergences \citep[Chapter 4]{pettigrew2016alc}.\footnote{If we use Bregman divergences to measure the distance from the omniscient credence function to another credence function, the resulting measure of inaccuracy is a \emph{strictly proper scoring rule}. These measures of inaccuracy have been justified independently in the accuracy-first literature \citep{oddie1997cccv, gibbard2007rcvt, joyce2009ac}. And conversely, given a strictly proper scoring rule, we can easily recover an additive Bregman divergence.} I won't rehearse the argument here, but I will accept the conclusion.

Now, on its own, my argument that only the additive Bregman divergences are legitimate for the purpose of measuring inaccuracy does not entail that only the additive Bregman divergences are legitimate for the purpose of correcting incoherent credences. But the following argument gives us reason to take that further step as well. One of the appealing features of the so-called accuracy-first approach to the epistemology of credences is that it gives a neat and compelling argument for the credal norm of \emph{probabilism}, which says that an agent should have a coherent credence function \citep{joyce1998nvp, pettigrew2016alc}. Having justified the restriction to the additive Bregman divergences on other grounds, the accuracy-first argument for probabilism is based on the following mathematical fact:
\begin{theorem}[\citep{predd2009pc}]\label{predd}
Suppose $\D$ is an additive Bregman divergence. And suppose $c$ is an incoherent credence function. Then, if $c^* = \underset{c' \in P_\F}{\mathrm{arg\, min}\ } \D(c', c)$, then $\D(v_i, c^*) < \D(v_i, c)$ for all $1 \leq i \leq m$, where $v_i(X_j) = 1$ if $i = j$ and $v_i(X_j) = 0$ if $i \neq j$.
\end{theorem}
That is, if $c$ is incoherent, then the closest coherent credence function to $c$ is closer to all the possible omniscient credence functions than $c$ is, and thus is more accurate than $c$ is at all possible worlds. Thus, if we fix up incoherent credence functions by using an additive Bregman divergence and taking the nearest coherent credence function, then we have an explanation for why we fix up incoherent credence functions in this way, namely, that doing so is guaranteed to increase the accuracy of the credence function. To see this in action, consider $\Fix_\SED(c_A)$ and $\Fix_\SED(c_B)$ in Figure \ref{amira-benito-sed}. It is clear from this picture that $\Fix_\SED(c_A)$ is closer to $v_X$ than $c_A$ is, and closer to $v_{\overline{X}}$ than $c_A$ is.

\section{The main results}\label{main-results}

\subsection{Minimizing distance from coherence}

From Proposition \ref{sed-gkl}(i) and (ii), we learned of an additive Bregman divergence that fixes up incoherent credences in a way that cooperates with linear pooling --- it is $\SED$. And we learned of an additive Bregman divergence that fixes up incoherent credences in a way that cooperates with geometric pooling, at least when you fix by minimizing distance \emph{from} coherence rather than distance \emph{to} coherence --- it is $\GKL$. But this leaves open whether there are other additive Bregman divergences that cooperate with either of these rules. The following theorem shows that there are not.
\begin{theorem}\label{sed-gkl-1-necessary}
Suppose $\D$ is an additive Bregman divergence. Then:
\begin{enumerate}
\item[\emph{(i)}]  $\WCAP_{\D_1} = \Fix_{\D_1} \circ \LP$ iff $\D$ is a positive linear transformation of $\SED$.
\item[\emph{(ii)}] $\WCAP_{\D_1} = \Fix_{\D_1} \circ \GP = \GP \circ \Fix_{\D_1}$ iff $\D$ is a positive linear transformation of $\GKL$.
\end{enumerate}
\end{theorem}
Thus, suppose you fix incoherent credences by minimizing distance \emph{from} coherence. And suppose you wish to fix and aggregate in ways that cooperate with one another  --- we will consider an argument for doing this in section \ref{phil-sig}. Then, if you measure the divergence between credence functions using $\SED$, then Proposition \ref{sed-gkl} says you should aggregate by linear pooling. If, on the other hand, you wish to use $\GKL$, then you should aggregate by geometric pooling. And, conversely, if you aggregate credences by linear pooling, then Theorem \ref{sed-gkl-1-necessary} says you should fix incoherent credences using $\SED$. If, on the other hand, you aggregate by geometric pooling, then you should fix incoherent credences using $\GKL$. In section \ref{phil-sig}, we will ask whether we have reason to fix and aggregate in ways that cooperate with one another.

We learned something further from Proposition \ref{sed-gkl}(i) and (ii).  We learned that $\SED$-aggregation commutes with $\SED$-fixing, and their combination agrees with $\WCAP_\SED$; and we learned that $\GKL_1$-aggregation commutes with $\GKL_1$-fixing, and their combination agrees with $\WCAP_{\GKL_1}$. The following result shows that this pattern generalises: $\D_1$-aggregation commutes with $\D_1$-fixing, and their combination agrees with $\WCAP_{\D_1}$, for any additive Bregman divergence $\D$.
\begin{theorem}\label{d-agg-wcap} Suppose $\D$ is an additive Bregman divergence. Then
\[
\WCAP_{\D_1} = \Fix_{\D_1} \circ \Agg_{\D_1} = \Agg_{\D_1} \circ \Fix_{\D_1}
\]
\end{theorem}
This is a useful result if you are already in possession of an additive Bregman divergence and you wish to find an aggregation method that will cooperate with it. 

We round off this section with a result that is unsurprising in the light of previous results:
\begin{theorem}\label{agg-unique-1} Suppose $\D$ is an additive Bregman divergence. Then,
\begin{enumerate}
\item[\emph{(i)}] $\Agg_{\D_1} = \LP$ iff $\D$ is a positive linear transformation of $\SED$.
\item[\emph{(ii)}] $\Agg_{\D_1} = \GP_-$ iff $\D$ is a positive linear transformation of $\GKL$.
\end{enumerate}
\end{theorem}

\subsection{Minimizing distance to coherence}

Next, let us consider what happens when we fix incoherent credences by minimizing distance \emph{to} coherence rather than distance \emph{from} coherence. %Bear in mind, however, that the justification given in section \ref{bregman} for using additive Bregman divergences also motivates fixing incoherent credences by minimizing distance \emph{from} coherence.
\begin{theorem}\label{sed-gkl-2-necessary}
Suppose $\D$ is an additive Bregman divergence. Then,
\begin{enumerate}
%\item[\emph{(i)}] $\Agg_{\D_2} = \LP$, if $\varphi''(x) > 0$, for $0 \leq x \leq 1$.
\item[\emph{(i)}] $\WCAP_{\D_2} = \Fix_{\D_2} \circ \LP $.
\item[\emph{(ii)}] $\WCAP_{\D_2} = \Fix_{\D_2} \circ \LP = \LP \circ \Fix_{\D_2}$, \emph{when the methods are applied to coherent credences}. 
\item[\emph{(iii)}] $\WCAP_{\D_2} = \Fix_{\D_2} \circ \LP = \LP \circ \Fix_{\D_2}$ iff $\D$ is a positive linear transformation of $\SED$.
\end{enumerate}
\end{theorem}
Theorem \ref{sed-gkl-2-necessary}(iii) gives the analogue to Theorem \ref{sed-gkl-1-necessary}(i). Only $\SED$ cooperates with linear pooling, even when we fix by minimizing distance \emph{to} coherence. On the other hand, Theorem \ref{sed-gkl-2-necessary}(i) and (ii) entail that there is no analogue to Theorem \ref{sed-gkl-1-necessary}(ii). There is no additive Bregman divergence that cooperates with geometric pooling when we fix by minimizing distance \emph{to} coherence. That is, there is no additive Bregman divergence $\D$ such that $\WCAP_{\D_2} = \Fix_{\D_2} \circ \GP = \GP \circ \Fix_{\D_2}$. This result complicates our thesis from above that $\SED$ pairs naturally with linear pooling while $\GKL$ pairs naturally with geometric pooling. %We might amend that thesis now by saying that $\SED$ pairs naturally with linear pooling while $\GKL$ pairs naturally with geometric pooling \emph{when anything does}. Or we might appeal to the remark at the end of section \ref{bregman}, where it was noted that accuracy considerations militate in favour of fixing by minimizing distance \emph{from} coherence --- only then is the fixed credence function guaranteed to be more accurate than the original. As Theorem \ref{sed-gkl-1-necessary} makes clear, if we focus on fixing minimizing distance \emph{from} coherence, there is no need to include a caveat concerning $\GKL$ and geometric pooling.

We round off this section with the analogue of Theorem \ref{agg-unique-1}:
\begin{theorem}
\label{agg-unique-2} Suppose $\D$ is an additive Bregman divergence. Then, $\Agg_{\D_2} = \LP$.
\end{theorem}

%\subsection{$\D$-aggregation and linear and geometric pooling}

%We round off this section with a result that should be rather unsurprising in the light of our other results:
%\begin{theorem}\label{agg-unique} Suppose $\D$ is the additive Bregman divergence generated by $\varphi$. Then,
%\begin{enumerate}
%\item[\emph{(i)}] $\Agg_{\D_1} = \LP$ iff $\D$ is a positive linear transformation of $\SED$.
%\item[\emph{(ii)}] $\Agg_{\D_1} = \GP_-$ iff $\D$ is a positive linear transformation of $\GKL$.

%\end{enumerate}
%\end{theorem}

\section{The philosophical significance of the results}\label{phil-sig}

What is the philosophical upshot of the results that we have presented so far? I think they are best viewed as supplements that can be added to existing arguments. On their own, they do not support any particular philosophical conclusion. But, combined with an existing philosophical argument, they extend its conclusion significantly. They are, if you like, philosophical booster rockets.

There are two ways in which the results above might provide such argumentative boosts. First, if you think that the aggregate of a collection of credence functions should be the credence function that minimizes the weighted average divergence from or to those functions, then you might appeal to Proposition \ref{aggregate} or Theorems \ref{agg-unique-1} and \ref{agg-unique-2} either to move from a way of measuring divergence to a method of aggregation, or to move from an aggregation method to a favoured divergence. Thus, given an argument for linear pooling, and an argument that you should aggregate by minimizing weighted average distance from the aggregate to the agent, you might cite Theorem \ref{agg-unique-1}(i) and argue for measuring how far one credence function lies from another using $\SED$. Or, given an argument that aggregation is minimizing weighted average divergence to the agents, and an argument in favour of $\GKL$, you might cite Theorem \ref{agg-unique-1}(ii) and conclude further that you should aggregate by $\GP_-$. Throw in an argument that you should fix incoherent credence functions by minimizing distance from coherence and this gives an argument for $\GP$.

Second, if you think that the three possible ways of producing a single coherent credence function from a collection of possibly incoherent ones should cooperate --- that is, if you think that aggregate-then-fix, fix-then-aggregate, and the weighted coherent approximation principle should all give the same outputs when supplied with the same input --- then you might appeal to Theorems \ref{sed-gkl-1-necessary} and \ref{sed-gkl-2-necessary} to move from aggregation method to divergence, or vice versa. For instance, if you think we should fix by minimizing distance from coherence, you might use Theorem \ref{sed-gkl-1-necessary}(i) to boost an argument for linear pooling so that it becomes also an argument for $\SED$, or to boost an argument for geometric pooling to give an argument for $\GKL$. And so on.

We begin, in this section, by looking at the bases for these two sorts of argument. Then we consider the sorts of philosophical argument to which our boosts might be applied. That is, we ask what sorts of arguments we might give in favour of one divergence over another, or one aggregation method over another, or whether we should fix by minimizing distance to or from coherence.

%For instance, if I have an argument that speaks in favour of using $\SED$ to measure how far one credence function lies from another, then Propositions \ref{aggregate}(i) and \ref{sed-gkl}(i) allow me to extend that argument so that it also supports using linear pooling to aggregate credences. Similarly, if I have good reason to measure the divergence from one credence function to another using $\GKL$ and reason to fix incoherent credences by minimizing divergence \emph{to} coherence, then Propositions \ref{aggregate}(ii) and \ref{sed-gkl}(iv) give me good reason to aggregate by linear pooling. On the other hand, if I have reason to use $\GKL$ as my divergence and reason to fix incoherence by minimizing divergence \emph{from} coherence, I thereby have reason to aggregate using geometric pooling. And conversely any reason to favour one or other of linear or geometric pooling, together with a reason to fix by minimizing divergence either from or to coherence, gives us reason to favour one or other of $\SED$ or $\GKL$ as our divergence via Theorems \ref{sed-gkl-1-necessary} and \ref{sed-gkl-2-necessary}. In this section, we begin by looking at the foundations of these arguments, to see whether our results really can provide the philosophical boost that they promise. Then we turn to consider the sorts of arguments to which they might contribute that boost. That is, we ask what sorts of arguments we might give in favour of one divergence over another, or one aggregation method over another, or whether we should fix by minimizing distance to or from coherence.

\subsection{Aggregating as minimizing weighted average distance}

Why think that we should aggregate the credence functions of a group of agents by finding the single credence function from or to which the weighted average distance is minimal? There is a natural argument that appeals to a principle that is used elsewhere in Bayesian epistemology. Indeed, we have used it already in this paper in our brief justification for fixing incoherent crecences by minimizing distance from or to coherence. It is \emph{the principle of minimal mutilation}. The idea is this: when you are given a collection of credences that you know are flawed in some way, and from which you wish to extract a collection that is not flawed, you should pick the unflawed collection that involves the least possible change to the original flawed credences.

The principle of minimal mutilation is often used in arguments for credal updating rules. Suppose you have a prior credence function, and then you acquire new evidence. Since it is new evidence, your prior likely does not satisfy the constraints that your new evidence places on your credences. How are you to respond? Your prior is now seen to be flawed --- it violates a constraint imposed by your evidence --- so you wish to find credences that are not flawed in this way. A natural thought is this: you should move to the credence function that does satisfy those constraints and that involves the least possible change in your prior credences; in our terminology, you should move to the credence function whose distance from or to your prior amongst those that satisfies the constraints is minimal. This is the principle of minimal mutilation in action. And its application has lead to a number of arguments for various updating rules, such as Conditionalization, Jeffrey Conditionalization, and others \citep{williams1980bcp, lewis1981cdt, diaconis1982usp, leitgeb2010ojb2, levinstein2012lpu}.

As we have seen in section \ref{fix-cred}, the principle of minimal mutilation is also our motivation for fixing an incoherent credence function $c$ by taking $\Fix_{\D_1}(c)$ or $\Fix_{\D_2}(c)$, for some divergence $\D$. And the same holds when you have a group of agents, each possibly incoherent, and some of whom disagree with each other. Here, again, the credences you receive are flawed in some way: within an individual agent's credence functions, the credences may not cohere with each other; and between agents, there will be conflicting credence assignments to the same proposition. We thus wish to find a set of credences that are not flawed in either of these ways. We want one credence per proposition, and we want all of the credences to cohere with one another. We do this by finding the set of such credences that involves as little change as possible from the original set. The weightings in the weighted average of the divergences allow us to choose which agent's credences we'd least like to change (they receive highest weighting) and whose we are happiest to change (they receive lowest weighting).

\subsection{The No Dilemmas argument}\label{no-dilemmas}
As we noted above, in order to use Theorem \ref{sed-gkl-1-necessary}(i), say, to extract a reason for using $\SED$ from a reason for aggregating by linear pooling, we must argue that the three possible ways of producing a single coherent credence function from a collection of possibly incoherent  credence functions \emph{should} cooperate. That is, we must claim that aggregate-then-fix, fix-then-aggregate, and the weighted coherent approximation principle \emph{should} all give the same outputs when supplied with the same input. The natural justification for this is a \emph{no dilemmas argument}. The point is that, if the three methods don't agree on their outputs when given the same set of inputs, we are forced to pick one of those different outputs to use. And if there is no principled reason to pick one or another, whichever we pick, we cannot justify using it rather than one of the others. Thus, for instance, given any decision where the different outputs recommend different courses of action, we cannot justify picking the action recommended by one of the outputs over the action recommended by one of the others. Similarly, given any piece of statistical reasoning in which using the different outputs as prior probabilities results in different conclusions at the end, we cannot justify adopting the conclusion mandated by one of the outputs over the conclusion mandated by one of the others.

Does this no dilemmas argument work? Of course, you might object if you think that there are principled reasons for preferring one method to another. That is, you might answer the no dilemmas argument by claiming that there is no dilemma in the first place, because one of the options is superior to the others. For instance, you might claim that it is more natural to fix first and then aggregate than to aggregate first and then fix. You might say that we can only expect an aggregate to be epistemically valuable when the credences to be aggregated are epistemically valuable; and you might go on to say that credences aren't epistemically valuable if they're incoherent.\footnote{Thanks to XX for urging me to address this line of objection.} But this claim is compatible with aggregating first and then fixing. I can still say that aggregates are only as epistemically valuable as the credence functions they aggregate, and I can still say that the more coherent a credence function the more epistemically valuable it is, and yet also say that I should aggregate and then fix. After all, while the aggregate won't be very epistemically valuable when the agents are incoherent, once I've fixed it and made it coherent it will be. And there's no reason to think it will be epistemically worse than if I first fixed the agents and then aggregated them. So I think this particular line of argument fails.

Here's another. There are many different reasons why an agent might fail to live up to the ideal of full coherence: the computations required to maintain coherence might be beyond their cognitive powers; or coherence might not serve a sufficiently useful practical goal to justify devoting the agent's limited cognitive resources to its pursuit; or an agent with credences over a partition might only ever have considered each cell of that partition on its own, separately, and never have considered the logical relations between them, and this might have lead her inadvertently to assign incoherent credences to them. So it might be that, while there is no reason to favour aggregating-then-fixing over fixing-then-aggregating or the weighted coherent approximation principle \emph{in general}, there is reason to favour one or other of these methods \emph{once we identify the root cause of the agent's incoherence}.

For instance, you might think that, when her incoherence results from a lack of attention to the logical relations between the propositions, it would be better to treat the individual credences in the individual members of the partition separately for as long as possible, since they were set separately by the agent. And this tells in favour of aggregating via $\LP$ or $\GP_-$ first, since the aggregate credence each sets in a given proposition is a function only of the credences that the agents assign to that proposition. I don't find this argument compelling. After all, it is precisely the fact that the agent has considered these propositions separately that has given rise to their flaw. Had they considered them together as members of one partition, they may come closer to the ideal of coherence. So it seems strange to wish to maintain that separation for as long as possible. It seems just as good to fix the flaw that has resulted from keeping them separate so far, and then aggregate the results. However, while I find the argument weak, it does show how we might look to the reasons behind the incoherence in a group of agents, or perhaps the reasons behind their disagreements, in order to break the dilemma and argue that the three methods for fixing and aggregating need not agree.

\subsection{Minimizing divergence from or to coherence}

As we have seen in Propositions \ref{aggregate} and \ref{sed-gkl} and Theorems \ref{sed-gkl-1-necessary} and \ref{sed-gkl-2-necessary}, it makes a substantial difference whether you fix incoherent credence functions by minimizing distance from or to coherence, and whether you aggregate credences by minimizing distance from or to the agents' credence functions when you aggregate them. Do we have reason to favour one of these directions or the other?

Here is one argument, at least in the case of fixing incoherent credences. Recall Theorem \ref{predd} from above. Suppose $c$ is an incoherent credence function. Then let $c^*$ be the coherent credence function for which the divergence \emph{from} $c^*$ \emph{to} $c$ is minimal, and let $c^\dag$ be the coherent credence function for which the distance \emph{to} $c^\dag$ \emph{from} $c$ is minimal. Then $c^*$ is guaranteed to be more accurate than $c$, while $c^\dag$ is not. Now, this gives us a reason for fixing an incoherent credence function by minimizing the distance \emph{from} coherence rather than the distance \emph{to} coherence. It explains why we should use $\Fix_{\D_1}$ rather than $\Fix_{\D_2}$ to fix incoherent credence functions. After all, when $\D$ is a Bregman divergence, $\Fix_{\D_1}(c)$ is guaranteed to be more accurate than $c$, by Theorem \ref{predd}, whereas $\Fix_{\D_2}(c)$ is not.

\subsection{Linear pooling vs geometric pooling}

In this section, we briefly survey some of the arguments for and against linear or geometric pooling. For useful surveys of the virtues and vices of different aggregation methods, see \citep{genest1986cpd, dietrich2015pop, russell2015g}.

In favour of aggregating by linear pooling ($\LP$): First, if each of the agents is coherent, then so is the aggregate. Having said that, given that, for many aggregation methods, Theorem \ref{d-agg-wcap} gives us ways to fix incoherent credences that commute with those methods, this isn't a practical desideratum. If we have an aggregation method, like $\GP_-$, that does not necessarily take coherent credences to a coherent aggregate, we can simply apply the relevant method for fixing incoherence --- in this case, $\Fix_\GKL$. 

Second, \cite{mcconway1981mlop} and \citet{wagner1982alm} show that, amongst the aggregation methods that always take coherent credence functions to coherent aggregates, linear pooling is the only one that satisfies what \citet{dietrich2015pop} call \emph{eventwise independence} and \emph{unanimity preservation}. Eventwise Independence demands that aggregation is done proposition-wise using the same method for each proposition. That is, an aggregation methods $T$ satisfies strong label neutrality if there is a function $f : [0, 1]^n \rightarrow [0, 1]$ such that $T(c_1, \ldots, c_n)(X_j) = f(c_1(X_j), \ldots, c_n(X_j))$ for each cell $X_j$ in our partition $\F$. Unanimity Preservation demands that, when all agents have the same credence function, their aggregate should be that credence function. That is, $T(c, \ldots, c) = c$, for any coherent credence function $c$. It is worth noting, however, that $\GP_-$ also satisfies both of these constraints; but of course it doesn't always take coherent credences to coherent aggregates.

Third, in a previous paper, I showed that linear pooling is recommended by the accuracy-first approach in epistemology, which we met in section \ref{bregman} \citep{pettigrewtaagc}. Suppose, like nearly all parties to the accuracy-first debate, you measure the accuracy of credences using what is known as a strictly proper scoring rule; this is equivalent to measuring the accuracy of a credence function at a world as the divergence from the omniscient credence function at that world to the credence function, where the divergence in question is an additive Bregman divergence. Suppose further that each of the credence functions you wish to aggregate is coherent. Then, if you aggregate by anything other than linear pooling, there will be an alternative aggregate credence function that each of the agents expects to be more accurate than your aggregate. I argue that a credence function cannot count as the aggregate of a set of credence functions if there is some alternative that each of those credence functions expects to do better epistemically speaking.

Against linear pooling: First, \citet{dalkey1975} notes that it does not commute with conditionalization.\footnote{For responses to this objection to linear pooling, see \citep{madansky1964am, mcconway1981mlop, pettigrewtaagc}.} Thus, if you first conditionalize your agents on a piece of evidence and then linear pool, this usually gives a different result from linear pooling first and then conditionalizing (at least if you use the same weights before and after the evidence is accommodated). That is, typically,
\[
\LP^{\{\alpha\}}(c_1, \ldots, c_n)(-|E) \neq \LP^{\{\alpha\}}(c_1(-|E), \ldots, c_n(-|E))
\]

Second, \citet{laddaga1977lcp} and \citet{lehrer1983pa} note that linear pooling does not preserve relationships of probabilistic independence.\footnote{For responses to this objection to linear pooling, see \citep{genest1987fe, wagner2010pd, pettigrewtaagc}.} Thus, usually, if $A$ and $B$ are probabilistically independent relative to each $c_i$, they will not be probabilistically independent relative to the linear pool. That is, if $c_i(A|B) = c_i(A)$ for each $c_i$, then usually
\[
\LP^{\{\alpha\}}(c_1, \ldots, c_n)(A|B) \neq \LP^{\{\alpha\}}(c_1, \ldots, c_n)(A).
\]

Geometric pooling ($\GP$) succeeds where linear pooling fails, and fails where linear pooling succeeds. The accuracy-first argument tells against it; and it violates Eventwise Independence. But it commutes with conditionalization. What's more, while linear pooling typically returns an incoherent aggregate when given incoherent agents, geometric pooling always returns a coherent aggregate, whether the agents are coherent or incoherent. Of course, this is because we build in that coherence by hand when we normalize the geometric averages of the agents' credences.

\subsection{Squared Euclidean distance vs generalized Kullback-Leibler divergence}

%While the results of the previous section might have some mathematical interest, and might be useful to those who are actually faced with a group of incoherent agents they wish to aggregate, it is less clear that they have philosophical significance. Nonetheless, I think they do.  I think they allow those with good antecedent reason to use one or other of $\SED$ or $\GKL$ as a divergence to choose between linear and geometric pooling as a method of aggregation. Those with reason to use $\SED$ are, via Theorem \ref{sed-gkl-1-necessary}(i), given reason to aggregate by linear pooling, while those with reason to use $\GKL$ are, via Theorem \ref{sed-gkl-1-necessary}(ii), given reason to pool geometrically. And, conversely, those with antecedent reason to pool linearly are given reason to measure divergence from one credence function to another using $\SED$, while those with reason to pool geometrically should opt for $\GKL$. And indeed there are plenty of ways in which one might acquire antecedent reasons to favour one divergence over another, or one method of aggregation over another.
There are a number of different ways in which we might argue in favour of $\SED$ or $\GKL$. 

In favour of squared Euclidean distance ($\SED$): First, there is an argument that I have offered elsewhere that proceeds in two steps \citep[Chapter 4]{pettigrew2016alc}: (i) we should measure how far one credence function lies from another using additive Bregman divergences, because only by doing so can we capture two competing senses of accuracy --- the alethic and the calibrationist --- in one measure; (ii) the distance from one credence function to another should be the same as the distance to the first credence from the second, so that our divergence should be symmetric. Since $\SED$ is the only symmetric Bregman divergence, this gives an argument in its favour.

Second, \citet{dagostino2010easp} argue for $\SED$ axiomatically. $\SED$ is the only way of measuring how far one credence function lies from another that satisfies certain plausible formal constraints. \citet{csiszar1991lsme, csiszar2008acim} offers axiomatic characterizations of $\SED$ and $\GKL$ that allow us to tell between on the basis of their formal features.

Third, some argue in favour of $\SED$ indirectly. They argue primarily in favour of the so-called Brier score.\footnote{According to the Brier score, the inaccuracy of a credence function $c$ at a world $w$ is 
\[
\B(c, w) = \sum^m_{i=1} (v_w(X_i) - c(X_i))^2
\]
where $v_w$ is the omniscient credence function at world $w$.} This is a particular inaccuracy measure that is widely used in the accuracy-first epistemology literature. It is a strictly proper scoring rule. And it is the inaccuracy measure that you obtain by using $\SED$ and taking inaccuracy to be divergence from omniscient credences. Thus, arguments for the Brier score can be extended to give arguments for $\SED$. How might we argue for the Brier score? First, \citet{schervish1989gm} showed that agents with different practical ends and different opinions about what decisions they are likely to face will value their credences for pragmatic purposes using different strictly proper scoring rules. Thus, we might argue for the Brier score if we have particular practical ends and if we hold a certain view about the sorts of decisions we'll be asked to make \citep{levinsteintapgeu}. Second, you might argue for the Brier score because of the way that it scores particular credences. It is more forgiving of extreme inaccuracy than is, for instance, the logarithmic scoring rule associated with $\GKL$ \citep{joyce2009ac}.

Fourth, as we will see in section \ref{partition}, when we consider credence functions defined over sets of propositions that don't form partitions, a dilemma arises when we use $\GKL$ that does not arise when we use $\SED$. As I will argue there, this speaks in favour of $\SED$.

Against $\SED$: if we use it to say how we should update in response to a certain sort of evidence, it gives updating rules that seem defective \citep{leitgeb2010ojb2, levinstein2012lp}. It does not justify either Bayesian Conditionalization or Jeffrey Conditionalization; and the alternative rules that it offers have undesirable features.

This argument against $\SED$ is also the primary argument in favour of $\GKL$. As I show elsewhere, it is difficult to find a Bregman divergence other than $\GKL$ that warrants updating by Conditionalization and Jeffrey Conditionalization \cite[Theorem 15.1.4]{pettigrew2016alc}.

%Some argue that we should use $\SED$ in accuracy-first epistemology because of its formal features \citep{leitgeb2010ojb1,pettigrew2016alc}. \citet{leitgeb2010ojb1} appeals to a no dilemmas quite different from the one we met above: they argue that there are different ways of measuring the inaccuracy of a credence functionOthers argue instead for $\GKL$ in that context because of the updating rules that it mandates . 

%Some favour linear pooling because it alone satisfies strong label neutrality \citep{wagner1982alm}. Others favour it because, if we use any other aggregation method, there is an alternative aggregate that all agents expect to be more accurate \citep{pettigrewtaagc}. And some favour geometric pooling because it commutes with conditionalization \citep{russell2015g}.

 %And while it is difficult to prove a negative existential claim, I see little prospect of finding any compelling reason to favour one of the three methods over the other two. So I submit that we should pick our divergence and our aggregation procedure so that the three methods for producing a single coherent credence function from a set of possibly incoherent credence functions agree. And this allows us to use Theorems \ref{sed-gkl-1-necessary} and \ref{sed-gkl-2-necessary} to turn reasons for using a particular aggregation procedure into reasons for using a particular divergence, and vice versa.

\section{Beyond partitions}\label{partition}

So far, we have restricted attention to credence functions defined on partitions. In this section, we lift that restriction.
%I have been neutral between linear pooling and geometric pooling, on the one hand, and $\SED$ and $\GKL$, on the other. In the previous section, I suggested that you should plump either for $\SED$ and linear pooling, or for $\GKL$ and geometric pooling. But I said nothing in favour of one of these packages rather than the other. In this section, however, I suggest that, when we drop the assumption that the credence functions are defined over a partition, we see that there are reasons to favour $\SED$ and linear pooling over $\GKL$ and geometric pooling. While the problem I will raise for $\GKL$ and geometric pooling arises from thinking about how to aggregate incoherent agents, we can state it entirely in terms of coherent agents.
Suppose Carmen and Donal are two further expert epidemiologists. They have credences in a rather broader range of propositions than Amira and Benito do. They consider the proposition, $X_1$, that the next `flu pandemic will occur in 2019, but also the proposition, $X_2$, that it will occur in 2020, the proposition $X_3$ that it will occur in neither 2019 nor 2020, and the proposition, $X_1 \vee X_2$, that it will occur in 2019 or 2020. Thus, they have credences in $X_1$, $X_2$, $X_3$, and $X_1 \vee X_2$, where the first three propositions form a partition but the whole set of four does not. Unlike Amira and Benito, Carmen and Donal are coherent. Here are their credences:
\begin{center}
\begin{tabular}{r|cccc}
& $X_1$ & $X_2$ & $X_3$ & $X_1 \vee X_2$  \\
\hline 
Carmen ($c_1$) &20\% & 30\% & 50\% & 50\% \\
Donal ($c_2$) & 60\% & 30\% & 10\% & 90\%
\end{tabular}
\end{center}
Since they are coherent, the question of how to fix them does not arise. So we are interested here only in how to aggregate them. If we opt to combine $\SED$ and linear pooling, there are three methods:
\begin{enumerate}
\item[(LP1)] Apply the method of linear pooling to the most fine-grained partition, namely, $X_1$, $X_2$, $X_3$, to give the aggregate credences for those three propositions. Then take the aggregate credence for $X_1 \vee X_2$ to be the sum of the aggregate credences for $X_1$ and $X_2$, as demanded by the axioms of the probability calculus.

For instance, suppose $\alpha = \frac{1}{2}$. Then
\begin{itemize}
\item $c^*(X_1) = \frac{1}{2}0.2 + \frac{1}{2}0.6 = 0.4$
\item $c^*(X_2) = \frac{1}{2}0.3 + \frac{1}{2}0.3 = 0.3$
\item $c^*(X_3) = \frac{1}{2}0.5 + \frac{1}{2}0.1 = 0.3$
\item $c^*(X_1 \vee X_2) =  c^*(X_1) + c^*(X_2) = 0.4 + 0.3 = 0.7$.
\end{itemize}
\item[(LP2)] Extend the method of linear pooling from partitions to more general sets of propositions in the natural way: the aggregate credence for a proposition is just the weighted arithmetic average of the  credences for that proposition.

Again, suppose $\alpha = \frac{1}{2}$. Then
\begin{itemize}
\item $c^*(X_1) = \frac{1}{2}0.2 + \frac{1}{2}0.6 = 0.4$
\item $c^*(X_2) = \frac{1}{2}0.3 + \frac{1}{2}0.3 = 0.3$
\item $c^*(X_3) = \frac{1}{2}0.5 + \frac{1}{2}0.1 = 0.3$
\item $c^*(X_1 \vee X_2) =  \frac{1}{2}0.5 + \frac{1}{2}0.9 = 0.7$.
\end{itemize}
\item[(LP3)] Apply $\WCAP_\SED$, so that the aggregate credence function is the coherent credence function that minimizes the arithmetic average of the squared Euclidean distances to the  credence functions.

Again, suppose $\alpha = \frac{1}{2}$. Then
\[
\underset{c' \in P_\F}{\mathrm{arg\, min}\ } \frac{1}{2}\SED(c', c_1) + \frac{1}{2}\SED(c', c_2) = \frac{1}{2} c_1 + \frac{1}{2} c_2.
\]
\end{enumerate}
It is easy to see that these three methods agree. And they continue to agree for any number of agents, any weightings, and any set of propositions. Does the same happen if we opt to combine $\GKL$ and geometric pooling? Unfortunately not. Here are the analogous three methods:
\begin{enumerate}
\item[(GP1)] Apply the method of geometric pooling to the most fine-grained partition, namely, $X_1$, $X_2$, $X_3$, to give the aggregate credences for those three propositions. Then take the aggregate credence for $X_1 \vee X_2$ to be the sum of the aggregate credences for $X_1$ and $X_2$, as demanded by the axioms of the probability calculus.

Suppose $\alpha = \frac{1}{2}$. Then
\begin{itemize}
\item $c^*(X_1) = \frac{\sqrt{0.2}\sqrt{0.6}}{\sqrt{0.2}\sqrt{0.6} + \sqrt{0.3}\sqrt{0.3} + \sqrt{0.5}\sqrt{0.1}} \approx 0.398$
\item $c^*(X_2) = \frac{\sqrt{0.3}\sqrt{0.3}}{\sqrt{0.2}\sqrt{0.6} + \sqrt{0.3}\sqrt{0.3} + \sqrt{0.5}\sqrt{0.1}} \approx 0.345$
\item $c^*(X_3)= \frac{\sqrt{0.5}\sqrt{0.1}}{\sqrt{0.2}\sqrt{0.6} + \sqrt{0.3}\sqrt{0.3} + \sqrt{0.5}\sqrt{0.1}}\approx 0.257$
\item $c^*(X_1 \vee X_2) =  c^*(X_1) + c^*(X_2) \approx 0.743$
\end{itemize}
\item[(GP2)] Extend the method of geometric pooling from partitions to more general sets of credence functions.

The problem with this method is that it isn't clear how to effect this extension. After all, when we geometrically pool credences over a partition, we start by taking weighted geometric averages and then we normalize. We can, of course, still take weighted geometric averages when we extend beyond partitions. But it isn't clear how we would normalize. In the partition case, we take a cell of the partition, take the weighted geometric average of the  credences in that cell, then divide through by the sum of the weighted geometric averages of the  credences in the various cells of the partition. But suppose that we try this once we add $X_1 \vee X_2$ to our partition $X_1$, $X_2$, $X_3$. The problem is that the normalized version of the weighted geometric average of the agents' credences in $X_1 \vee X_2$ is not the sum of the normalized versions of the weighted geometric averages of the  credences in $X_1$ and in $X_2$. But how else are we to normalize?
\item[(GP3)] Apply $\WCAP_\SED$, so that the aggregate credence function is the coherent credence function that minimizes the arithmetic average of the generalized Kullback-Leibler divergence from that credence function to the  credence functions.

Again, suppose $\alpha = \frac{1}{2}$. Now, we can show that, if $c^* = \underset{c' \in P_\F}{\mathrm{arg\, min}\ } \frac{1}{2}\GKL(c', c_1) + \frac{1}{2}\GKL(c', c_2)$, then
\begin{itemize}
\item $c^*(X_1) = 0.390$
\item $c^*(X_2) = 0.338$
\item $c^*(X_3) = 0.272$
\item $c^*(X_1 \vee X_2) = 0.728$
\end{itemize}
\end{enumerate}
Thus, (GP2) does not work --- we cannot formulate it. And (GP1) and (GP3) disagree. This creates a dilemma for those who opt for the package containing $\GKL$ and geometric pooling. How should they aggregate  credences when the agents have credences in propositions that don't form a partition? Do they choose (GP1) or (GP3)? To avoid the dilemma, I claim, we should opt for $\SED$ and linear pooling instead.

\section{Conclusion}\label{conclusion}

This completes our investigation into the methods by which we might produce a single coherent credence function from a group of possibly incoherent expert credence functions. At the heart of our investigation is a set of results that suggest that squared Euclidean distance pairs naturally with linear pooling, while the generalized Kullback-Leibler divergence pairs naturally with geometric pooling. I suggested that these results might be used by philosophers to argue for an aggregation method if they have reason to favour a particular divergence, or to argue for a particular divergence if they have reason to favour one aggregation method over another.

\section{Appendix: Proofs}\label{proofs}

\subsection{Two useful lemmas}

We begin by stating two useful lemmas to which we will appeal in our proofs. Throughout, we suppose:
\begin{itemize}
\item $\D$ is the additive Bregman divergence generated by $\varphi$;
\item $\alpha_1, \ldots, \alpha_n \geq 0$ and $\sum^n_{k= 1} \alpha_k = 1$;
\item $c$, $c'$ are credence functions defined on the partition $\F = \{X_1, \ldots, X_m\}$;
\item $C_\F = \{c :  \{X_1, \ldots, X_m\} \rightarrow [0, 1]\}$
\item $P_\F = \{c : \{X_1, \ldots, X_m\} \rightarrow [0, 1]\, |\, \sum^m_{i=1} c(X_i) = 1\}$.
\end{itemize}

\begin{lemma}\label{useful-lem} Suppose $\{\alpha\} = \{\alpha_1, \ldots, \alpha_k\}$ is a set of weights for the experts.
\begin{enumerate}
\item[\emph{(i)}] $c^* = \underset{c' \in C_\F}{\mathrm{arg\, min}\ } \sum^n_{k=1} \alpha_k \D(c', c_k)$ iff
\[
\varphi'(c^*(X_j)) =  \sum^n_{k=1} \alpha_k \varphi'(c_k(X_j))
\]
for all $1 \leq j \leq m$.
\item[\emph{(ii)}] $c^* = \underset{c' \in P_\F}{\mathrm{arg\, min}\ } \sum^n_{k=1} \alpha_k \D(c', c_k)$ iff $c^*$ is coherent and there is a constant $K$ such that
\[
\varphi'(c^*(X_j)) - \sum^n_{k=1} \alpha_k  \varphi'(c_k(X_j))  = K
\]
for all $1 \leq j \leq m$.
\end{enumerate}
\end{lemma}
\emph{Proof of Lemma \ref{useful-lem}. }
\begin{enumerate}
\item[(i)] Consider the function
\begin{eqnarray*}
& & \sum^n_{k=1} \alpha_k \D\left ((x_1, \ldots, x_m), (c_k(X_1), \ldots, c_k(X_m))\right ) \\
& =&  \sum^n_{k=1} \alpha_k\left ( \sum^{m}_{i=1}   \left ( \varphi(x_i) - \varphi(c_k(X_i)) - \varphi'(c_k(X_i))(x_i - c_k(X_i)) \right )\right) \\
& =&   \sum^{m}_{i=1} \sum^n_{k=1} \alpha_k  \left ( \varphi(x_i) - \varphi(c_k(X_i)) - \varphi'(c_k(X_i))(x_i - c_k(X_i)) \right )
\end{eqnarray*}
Then, for $1 \leq j \leq m$,
\begin{eqnarray*}
& & \frac{\partial}{\partial x_j} \sum^n_{k=1} \alpha_k \D\left ((x_1, \ldots, x_m), (c_k(X_1), \ldots, c_k(X_m))\right ) \\
& = & \sum^n_{k=1} \alpha_k  \left (\varphi'(x_j) - \varphi'(c_k(X_j)) \right ) \\
& = &  \varphi'(x_j) - \sum^n_{k=1} \alpha_k  \varphi'(c_k(X_i))
\end{eqnarray*}
as required.
\item[(ii)] Consider the function
\begin{eqnarray*}
& & \sum^n_{k=1} \alpha_k \D\left ((x_1, \ldots, x_{m-1}, 1 - \sum^{m-1}_{l = 1} x_l), (c_k(X_1), \ldots, c_k(X_m))\right ) \\
& =&  \sum^n_{k=1} \alpha_k \left ( \sum^{m-1}_{i=1}  \left ( \varphi(x_i) - \varphi(c_k(X_i)) - \varphi'(c_k(X_i))(x_i - c_k(X_i)) \right ) + \right.\\
& & \left.  \left ( \varphi(1 - \sum^{m-1}_{l = 1} x_l) - \varphi(c_k(X_m)) - \varphi'(c_k(X_m))(1 - \sum^{m-1}_{l = 1} x_l - c_k(X_m)) \right ) \right )
\end{eqnarray*}
Then, for $1 \leq j \leq m-1$,
\begin{eqnarray*}
& & \frac{\partial}{\partial x_j} \sum^n_{k=1} \alpha_k \D\left ((x_1, \ldots, x_{m-1}, 1 - \sum^{m-1}_{l = 1} x_l), (c_k(X_1), \ldots, c_k(X_m))\right ) \\
& = & \sum^n_{k=1} \alpha_k  \left (\varphi'(x_j) - \varphi'(c_k(X_j)) \right ) - \sum^n_{k=1} \alpha_k  \left (\varphi'(1 - \sum^{m-1}_{l = 1} x_l) - \varphi'(c_k(X_m)) \right ) 
\end{eqnarray*}
%iff, $\sum^m_{i=1} x_i = 1$ and there is a constant $K$ such that, for $1 \leq i \leq m$,
%\[
%\sum^n_{k=1} \alpha_k  \left (\varphi'(x_i) - \varphi'(c_k(X_i)) \right ) = K
%\]
as required. 
\end{enumerate}

\begin{lemma}\label{useful-lem-2} Suppose $\{\alpha\}$ is a set of weights for the experts. Then
\begin{enumerate}
\item[\emph{(i)}]  $c^* = \underset{c' \in C_\F}{\mathrm{arg\, min}\ } \sum^n_{k=1} \alpha_k \D(c_k, c')$ iff
\[
\left (c^*(X_j) - \sum^n_{k=1} \alpha_k c_k(X_j) \right ) \varphi''(c^*(X_j)) =0
\]
for all $1 \leq j \leq m$.
\item[\emph{(ii)}]  $c^* = \underset{c' \in P_\F}{\mathrm{arg\, min}\ } \sum^n_{k=1} \alpha_k \D(c_k, c')$ iff $c^*$ is coherent and there is a constant $K$ such that
\[
\left (c^*(X_j) - \sum^n_{k=1} \alpha_k c_k(X_j) \right ) \varphi''(c^*(X_j)) = K
\]
for all $1 \leq j \leq m$.
\end{enumerate}
\end{lemma}
\emph{Proof of Lemma \ref{useful-lem-2}. }
\begin{enumerate}
\item[(i)] Consider the function
\begin{eqnarray*}
& & \sum^n_{k=1} \alpha_k \D\left ((c_k(X_1), \ldots, c_k(X_m)), (x_1, \ldots, x_m)\right) \\
& =&  \sum^n_{k=1} \alpha_k \left ( \sum^{m}_{i=1} \varphi(c_k(X_i)) - \varphi(x_i) - \varphi'(x_i)(c_k(X_i) - x_i) \right )
\end{eqnarray*}
Then, for $1 \leq j \leq m$,
\begin{eqnarray*}
& & \frac{\partial}{\partial x_j} \sum^n_{k=1} \alpha_k \D\left ((c_k(X_1), \ldots, c_k(X_m)), (x_1, \ldots, x_m)\right ) \\
& = & \left (x_j - \sum^n_{k=1} \alpha_k c_k(X_j) \right ) \varphi''(x_j)
\end{eqnarray*}
as required.
\item[(ii)] Consider the function
\begin{eqnarray*}
& & \sum^n_{k=1} \alpha_k \D\left ((c_k(X_1), \ldots, c_k(X_m)), (x_1, \ldots, x_{m-1}, 1 - \sum^{m-1}_{l = 1} x_l)\right) \\
& =&  \sum^n_{k=1} \alpha_k \left ( \sum^{m-1}_{i=1} \varphi(c_k(X_i)) - \varphi(x_i) - \varphi'(x_i)(c_k(X_i) - x_i) + \right.\\
& & \left. \varphi(c_k(X_m))  - \varphi(1 - \sum^{m-1}_{l = 1} x_l) - \varphi'(1 - \sum^{m-1}_{l = 1} x_l)(c_k(X_m) - 1 + \sum^{m-1}_{l = 1} x_l) \right )
\end{eqnarray*}
Then, for $1 \leq j \leq m$,
\begin{eqnarray*}
& & \frac{\partial}{\partial x_j} \sum^n_{k=1} \alpha_k \D\left ((c_k(X_1), \ldots, c_k(X_m)), (x_1, \ldots, x_{m-1}, 1 - \sum^{m-1}_{l = 1} x_l)\right ) \\
& = & \left (x_j - \sum^n_{k=1} \alpha_k c_k(X_j) \right ) \varphi''(x_j) - \left (1 - \sum^{m-1}_{l = 1} x_l - \sum^n_{k=1} \alpha_k c_k(X_m) \right ) \varphi''(1 - \sum^{m-1}_{l = 1} x_l) \\
\end{eqnarray*}
as required.
%Thus,
%\[
%\frac{\partial}{\partial x_i} \sum^n_{k=1} \alpha_k \D((x_1, \ldots, x_{m-1}, 1 - \sum^{m-1}_{l = 1} x_l), (c_{k1}, \ldots, c_{km})) = 0
%\]
%iff 
%\[
%(x_i - \sum^n_{k=1} \alpha_k c_k(X_i)) \varphi''(x_i) = (x_j - \sum^n_{k=1} \alpha_k c_k(X_j)) \varphi''(x_j))
%\]
%Thus, $c^* = \underset{c' \in P_\F}{\mathrm{arg\, min}\ } \sum^n_{k=1} \alpha_k \D(c_k, c')$ iff
%\[
%(c^*(X_i) - \sum^n_{k=1} \alpha_k c_k(X_i)) \varphi''(c^*(X_i)) = (c^*(X_j) - \sum^n_{k=1} \alpha_k c_k(X_j)) \varphi''(c^*(X_j))
%\]
%for all $1 \leq i, j \leq m$.
\end{enumerate}
This completes the proof. \hfill $\Box$

\begin{corollary}\label{useful-col} Suppose $\D$ is the additive Bregman divergence generated by $\varphi$. Then
\begin{enumerate}
\item[\emph{(i)}] $c^* = \underset{c' \in P_\F}{\mathrm{arg\, min}\ } \D(c', c)$ iff $c^*$ is coherent and there is a constant $K$ such that 
\[
\varphi'(c^*(X_j)) - \varphi'(c(X_j)) = K
\]
for all $1 \leq j \leq m$.
\item[\emph{(ii)}] $c^* = \underset{c' \in P_\F}{\mathrm{arg\, min}\ } \D(c, c')$ iff $c^*$ is coherent and there is a constant $K$ such that 
\[
(c^*(X_j) - c(X_j))\varphi''(c^*(X_j)) = K
\]
for all $1 \leq j \leq m$.
\end{enumerate}
\end{corollary}
\emph{Proof of Corollary \ref{useful-col}. } These follow immediately from Lemmas \ref{useful-lem}(ii) and \ref{useful-lem-2}(ii), respectively, when $n = 1$. \hfill $\Box$

\subsection{Proof of Proposition \ref{fix-sed-gkl}}

\emph{Proof of Proposition \ref{fix-sed-gkl}(i) } $\SED$ is the additive Bregman divergence generated by $\varphi(x) = x^2$. Thus, $\varphi'(x) = 2x$. What's more, $\SED$ is symmetric; so $\Fix_{\SED_1}(c) = \Fix_{\SED_2}(c)$. Thus, by Corollary \ref{useful-col}(i), $c^* = \Fix_{\SED_1}(c) = \Fix_{\SED_2}(c) = \underset{c' \in P_\F}{\mathrm{arg\, min}\ } \SED(c', c)$ iff $c^*$ is coherent and there is $K$ such that $2(c^*(X_j) - c(X_j)) = K$ (for all $1 \leq j \leq m$) iff $c^*(X_j) = c(X_j) + \frac{1-\sum^m_{i=1}c(X_i)}{m}$ (for all $1 \leq j \leq m$).\hfill $\Box$\bigskip

\noindent \emph{Proof of Proposition \ref{fix-sed-gkl}(ii) } $\GKL$ is the additive Bregman divergence generated by $\varphi(x) = x \log x - x$. Thus, $\varphi'(x) = \log x$. Thus, by Corollary \ref{useful-col}(i), $c^* = \Fix_{\GKL_1}(c) = \underset{c' \in P_\F}{\mathrm{arg\, min}\ } \GKL(c', c)$ iff $c^*$ is coherent and there is $K$ such that $\log c^*(X_j) - \log c(X_j) = K$ (for all $1 \leq j \leq m$) iff $c^*$ is coherent and there is $K$ such that $c^*(X_j) = K \cdot c(X_j)$ (for all $1 \leq j \leq m$) iff $c^*(X_j) =  \frac{c(X_j)}{\sum^m_{i=1} c(X_i)}$ (for all $1 \leq j \leq m$). \hfill $\Box$ \medskip

\noindent Also, by Corollary \ref{useful-col}(ii), $c^* = \Fix_{\GKL_2}(c) = \underset{c' \in P_\F}{\mathrm{arg\, min}\ } \GKL(c, c')$ iff $c^*$ is coherent and there is $K$ such that $\frac{c^*(X_j) - c(X_j)}{c^*(X_j)} = K$ (for all $1 \leq j \leq m$) iff $c^*$ is coherent and there is $K$ such that $\frac{c(X_j)}{c^*(X_j)} = K$ (for all $1 \leq j \leq m$) iff $c^*$ is coherent and there is $K$ such that $c^*(X_j) = K \cdot c(X_j)$ (for all $1 \leq j \leq m$) iff $c^*(X_j) =  \frac{c(X_j)}{\sum^m_{i=1} c(X_i)}$ (for all $1 \leq j \leq m$). \hfill $\Box$

\subsection{Proof of Proposition \ref{commutes}}

\noindent \emph{Proof of Proposition \ref{commutes}(i) }
By Proposition \ref{fix-sed-gkl}(i),
\[
\Fix_\SED(c_k)(X_j) = c(X_j) + \frac{1 - \sum^m_{i=1} c_k(X_i)}{m}
\]
So
\begin{eqnarray*}
& & \LP^{\{\alpha\}}(\Fix_\SED(c_1), \ldots, \Fix_\SED(c_n))(X_j) \\
& = & \sum^n_{k=1} \alpha_k \Fix_\SED(c_k)(X_j) \\
& = & \sum^n_{k=1} \alpha_k \left ( c_k(X_j) + \frac{1 - \sum^m_{i=1} c_k(X_i)}{m} \right ) \\
%& = & \sum^n_{i=1} \alpha_i c_i(X_j) + \sum^n_{i=1} \alpha_i\frac{1 - \sum^m_{k=1} c_i(X_k)}{m} \\
& = & \sum^n_{k=1} \alpha_k c_k(X_j) + \frac{1 - \sum^m_{i=1} \sum^n_{k=1} \alpha_k c_k(X_i)}{m}
\end{eqnarray*}
By Proposition \ref{fix-sed-gkl}(i),
\begin{eqnarray*}
& & \Fix_\SED(\LP^{\{\alpha\}}(c_1, \ldots, c_n)) \\
& = & \Fix_\SED(\sum^{n}_{k=1} \alpha_k c_k)(X_j) \\
& = & \sum^{n}_{k=1} \alpha_k c_k(X_j) + \frac{1 - \sum^m_{i=1} \sum^n_{k=1} \alpha_k c_k(X_i)}{m}
\end{eqnarray*}
Thus,
\[
 \LP^{\{\alpha\}}(\Fix_\SED(c_1), \ldots, \Fix_\SED(c_n))(X_j) = \Fix_\SED(\LP^{\{\alpha\}}(c_1, \ldots, c_n))(X_j)
\]
as required. \hfill $\Box$\bigskip

\noindent \emph{Proof of Proposition \ref{commutes}(iii) } By Proposition \ref{fix-sed-gkl}(ii),
\[
\Fix_\GKL(c_k)(X_j) = \frac{c_k(X_j)}{\sum^m_{i=1} c_k(X_i)}
\]
So,
\begin{eqnarray*}
& & \GP^{\{\alpha\}}(\Fix_\GKL(c_1), \ldots, \Fix_\GKL(c_n))(X_j) \\
& = & \frac{\prod^n_{k=1}\Fix_\GKL(c_k)(X_j)^{\alpha_k}}{\sum^m_{i=1} \prod^n_{k=1}\Fix_\GKL(c_k)(X_i)^{\alpha_k}} \\
& = & \frac{\prod^n_{k=1}\left ( \frac{c_k(X_j)}{\sum^m_{i=1} c_k(X_i)} \right )^{\alpha_k}}{\sum^m_{l=1} \prod^n_{k=1}\left (\frac{c_k(X_l)}{\sum^m_{i=1} c_k(X_i)} \right )^{\alpha_k}} \\
& = & \frac{\prod^n_{k=1}c_k(X_j)^{\alpha_k}}{\sum^m_{l=1} \prod^n_{k=1}c_k(X_l)^{\alpha_k}}
\end{eqnarray*}
Now, recall that, since geometric pooling always results in a coherent credence function,
\[
\Fix_\GKL(\GP^{\{\alpha\}}(c_1, \ldots, c_n)) = \GP^{\{\alpha\}}(c_1, \ldots, c_n) 
\]
And
\[
\GP^{\{\alpha\}}(c_1, \ldots, c_n))(X_j) = \frac{\prod^n_{k=1}c_k(X_j)^{\alpha_k}}{\sum^m_{l=1} \prod^n_{k=1}c_k(X_l)^{\alpha_k}}
\]
So
\[
\Fix_\GKL(\GP^{\{\alpha\}}(c_1, \ldots, c_n))(X_j) = \GP^{\{\alpha\}}(\Fix_\GKL(c_1), \ldots, \Fix_\GKL(c_n))(X_j) 
\]
as required. \hfill $\Box$

\subsection{Proof of Proposition \ref{aggregate} } 
This follows from Theorem \ref{agg-unique-1}, which we prove below.

\subsection{Proof of Proposition \ref{sed-gkl} } 

\emph{Proof of Propositions \ref{sed-gkl}(i) and (ii) } These are immediate consequences of Propositions \ref{commutes} and \ref{aggregate} and Theorem \ref{sed-gkl-1-necessary}. \hfill $\Box$\bigskip

%\noindent \emph{Proof of Proposition \ref{sed-gkl}(ii) } By Lemma \ref{useful-lem}(ii), $c^* = \WCAP^{\{\alpha\}}_{\GKL_1}(c_1, \ldots, c_n) = \underset{c' \in P_\F}{\mathrm{arg\, min}\ } \sum^n_{k=1} \alpha_k \GKL(c', c_k)$ iff $c^*$ is coherent and there is $K$ such that
%\[
%\log c^*(X_j) - \sum^n_{k=1} \alpha_k \log c_k(X_j) = K 
%\]
%(for all $1 \leq j \leq m$) iff $c^*$ is coherent and there is $K$ such that $c^*(X_j) = K\cdot \prod^n_{k=1} c_k(X_j)^{\alpha_k}$ (for all $1 \leq j \leq m$) iff
%\[
%c^*(X_j) = \frac{\prod^n_{k=1} c_k(X_j)^{\alpha_k}}{\sum^m_{i=1} \prod^n_{k=1} c_k(X_i)^{\alpha_k}}
%\]
%(for all $1 \leq j \leq m$) as required. \hfill $\Box$\bigskip

\noindent \emph{Proof of Proposition \ref{sed-gkl}(iii) } By Lemma \ref{useful-lem-2}(ii), $c^* = \WCAP^{\{\alpha\}}_{\GKL_2}(c_1, \ldots, c_n) = \underset{c' \in P_\F}{\mathrm{arg\, min}\ } \sum^n_{k=1} \alpha_k \GKL(c_k, c')$ iff $c^*$ is coherent and there is $K$ such that
\[
\frac{c^*(X_j) - \sum^n_{k=1} \alpha_k c_k(X_j)}{c^*(X_j)} = K
\]
(for all $1 \leq j \leq m$) iff $c^*$ is coherent and there is $K$ such that
\[
\frac{\sum^n_{k=1} \alpha_k c_k(X_j)}{c^*(X_j)} = K
\]
(for all $1 \leq j \leq m$) iff $c^*$ is coherent and there is $K$ such that $c^*(X_j) = K \cdot \sum^n_{k=1} \alpha_k c_k(X_j)$ (for all $1 \leq j \leq m$) iff 
\[
c^*(X_j) = \frac{\sum^n_{k=1} \alpha_k c_k(X_j)}{\sum^m_{i=1}\sum^n_{k=1} \alpha_k c_k(X_i)}
\]
(for all $1 \leq j \leq m$) as required. \hfill $\Box$\bigskip

\noindent \emph{Proof of Proposition \ref{sed-gkl}(iv) }
\[
\Agg^{\{\alpha\}}_{\GKL_2}(c_1, \ldots, c_n)(X_j) = \sum^n_{k=1} \alpha_k c_k(X_j)
\]
and
\[
\Fix_{\GKL_2}(c)(X_j) = \frac{c(X_j)}{\sum^m_{i=1} c(X_i)}
\]
So
\[
\Fix_{\GKL_2}(\Agg^{\{\alpha\}}_{\GKL_2}(c_1, \ldots, c_n))(X_j) = \frac{\sum^n_{k=1} \alpha_k c_k(X_j)}{\sum^m_{i=1} \sum^n_{k=1} \alpha_k c_k(X_i)} = \WCAP^{\{\alpha\}}_{\GKL_2}(c_1, \ldots, c_n)(X_j)
\]
as required. \hfill $\Box$

\subsection{Proof of Propositions \ref{dictator-1} and \ref{dictator-2} }

Recall:
\[
\WGCAP^{\{\alpha\}}_{\D_1}(c_1, \ldots, c_n) = \underset{c' \in P_\F}{\mathrm{arg\, min}\ } \prod^n_{k=1}\D(c', c_k)^{\alpha_k}
\]
First, suppose $c' \neq c_k$, for all $1 \leq k \leq n$. Then, by the definition of a divergence, for all $1 \leq k \leq n$, $\D(c', c_k) > 0$. Thus, $\prod^n_{k=1}\D(c', c_k)^{\alpha_k} > 0$. Next, suppose $c' = c_k$ for some $1 \leq k \leq n$. Then, again by the definition of a divergence, $\D(c', c_k) = 0$. Thus, $\prod^n_{k=1}\D(c', c_k)^{\alpha_k} = 0$. Thus, $\prod^n_{k=1}\D(c', c_k)^{\alpha_k}$ is minimized iff $c' = c_k$ for some $1 \leq k \leq n$. And similarly for $\WGCAP^{\{\alpha\}}_{\D_2}$, $\GAgg_{\D_1}$, and $\GAgg_{\D_2}$. \hfill $\Box$

\subsection{Proof of Theorem \ref{sed-gkl-1-necessary} } 

\noindent \emph{Proof of Theorem \ref{sed-gkl-1-necessary}(i) }  Suppose $\Fix_{\D_1} \circ \LP = \WCAP_{\D_1}$. Given $0 \leq a, b \leq 1$ and $0  \leq \alpha \leq 1$, let $(x^*, 1-x^*)$ be the coherent credence function that results from applying both procedures to the credence functions $(a, 0)$ (which assigns $a$ to $X$ and $0$ to $\overline{X}$) and $(b, 0)$ (which assigns $b$ to $X$ and $0$ to $\overline{X}$) --- it assigns $x^*$ to $X$ and $1-x^*$ to $\overline{X}$. That is,
\[
(x^*, 1-x^*) = \Fix_{\D_1}(\LP^\alpha((a, 0), (b, 0)) = \WCAP^\alpha_{\D_1}((a, 0), (b, 0))
\]
 By Corollary \ref{useful-col}(i),
\[
\varphi'(x^*) - \varphi'(1-x^*) = \varphi'(\alpha a + (1-\alpha) b) - \varphi'(\alpha \cdot 0 + (1-\alpha) \cdot 0)
\]
And by Lemma \ref{useful-lem}(ii),
\[
\varphi'(x^*) - \varphi'(1-x^*) = (\alpha \varphi'(a) + (1-\alpha) \varphi'(b)) - (\alpha \varphi'(0) + (1-\alpha) \varphi'(0))
\]
So
\[
\varphi'(\alpha a + (1-\alpha)b) - \varphi'(\alpha \cdot 0 + (1-\alpha) \cdot 0) = (\alpha \varphi'(a) + (1-\alpha) \varphi'(b)) - (\alpha \varphi'(0) + (1-\alpha) \varphi'(0))
\]
So
\[
\varphi'(\alpha a + (1-\alpha)b) = \alpha \varphi'(a) + (1-\alpha)\varphi'(b)
\]
Thus, $\varphi'(x) = kx + c$ for some constants $k, c$. And so $\varphi(x) = mx^2 + kx + c$, for some constants $m, k, c$. Since $\varphi$ is strictly convex, $m > 0$. Now, it turns out that, if $\psi$ is a strictly convex function and $\theta(x) = \psi(x) + kx + c$, then $\varphi$ and $\psi$ generate the same Bregman divergence. After all,
\begin{eqnarray*}
\theta(x) - \theta(y) - \theta'(y)(x - y) & = & (\psi(x) + kx + c) - (\psi(y) + ky + c) - (\psi'(x) + k)(x-y) \\
& = &\psi(x) - \psi(y) - \psi'(y)(x-y)
\end{eqnarray*}
So $\D$ is a positive linear transformation of $\SED$, as required. \hfill $\Box$\bigskip

\noindent \emph{Proof of Theorem \ref{sed-gkl-1-necessary}(ii) } Suppose $\Fix_{\D_1} \circ \GP = \GP \circ \Fix_{\D_1}  = \WCAP_{\D_1}$. First, apply $\Fix_{\D_1} \circ \GP$ to a single credence function $(a, b)$, which assigns $a$ to $X$ and $b$ to $\overline{X}$. Since geometric pooling also fixes, $\Fix_{\D_1} \circ \GP = \GP$, so they both take $(a, b)$ and return $\left (\frac{a}{a+b}, \frac{b}{a+b}\right)$. Since $\Fix_{\D_1} \circ \GP = \GP \circ \Fix_{\D_1}$, this is also the result of applying $\GP \circ \Fix_{\D_1}$ to $(a, b)$. Thus, by Corollary \ref{useful-col}(i), we have
\begin{equation}\label{identity1}
\varphi' \left (\frac{a}{a+b}\right ) - \varphi' \left (\frac{b}{a+b}\right ) = \varphi'(a) - \varphi'(b)
\end{equation}
for all $0 \leq a, b \leq 1$. We will use this identity below.\medskip

\noindent Next, since $\Fix_{\D_1} \circ \GP =  \WCAP_{\D_1}$, and since $\Fix_{\D_1} \circ \GP$ is just geometric pooling, and since geometric pooling takes two credence functions $(a, b)$ and $(a', b')$ and returns
\[
\left ( \frac{a^\alpha a'^{1-\alpha}}{a^\alpha a'^{1-\alpha} + b^\alpha b'^{1-\alpha}}, \frac{b^\alpha b'^{1-\alpha}}{a^\alpha a'^{1-\alpha} + b^\alpha b'^{1-\alpha}} \right ) 
\]
then $\WCAP_{\D_1}$ must return that too when given $(a, b)$ and $(a', b')$. Thus, by Lemma \ref{useful-lem}(ii),
\begin{eqnarray*}
& & \varphi'\left ( \frac{a^\alpha a'^{1-\alpha}}{a^\alpha a'^{1-\alpha} + b^\alpha b'^{1-\alpha}}\right ) - \varphi'\left ( \frac{b^\alpha b'^{1-\alpha}}{a^\alpha a'^{1-\alpha} + b^\alpha b'^{1-\alpha}}\right ) \\
& = & (\alpha \varphi'(a) + (1-\alpha) \varphi'(a')) - (\alpha \varphi'(b) + (1-\alpha)\varphi'(b'))
\end{eqnarray*}
Now, by the identity (\ref{identity1}) proved above, we have
\[
\varphi'\left ( \frac{a^\alpha a'^{1-\alpha}}{a^\alpha a'^{1-\alpha} + b^\alpha b'^{1-\alpha}}\right ) - \varphi'\left ( \frac{b^\alpha b'^{1-\alpha}}{a^\alpha a'^{1-\alpha} + b^\alpha b'^{1-\alpha}}\right ) = \varphi'(a^\alpha a'^{1-\alpha}) - \varphi'(b^\alpha b'^{1-\alpha})
\]
So
\begin{equation}\label{identity2}
\varphi'(a^\alpha a'^{1-\alpha}) - \varphi'(b^\alpha b'^{1-\alpha}) = (\alpha \varphi'(a) + (1-\alpha) \varphi'(a')) - (\alpha \varphi'(b) + (1-\alpha)\varphi'(b'))
\end{equation}
for all $0 \leq a, b, a', b' \leq 1$ and $0 \leq \alpha \leq 1$. So let $b = a' = b' = 1$. Then
\[
\varphi'(a^\alpha) - \varphi'(1) = (\alpha \varphi'(a) + (1-\alpha) \varphi'(1)) - (\alpha \varphi'(1) + (1-\alpha)\varphi'(1))
\]
So
\begin{equation}\label{identity3}
\varphi'(a^\alpha) = \alpha \varphi'(a) + (1-\alpha)\varphi'(1)
\end{equation}
for all $0 \leq a \leq 1$ and $0 \leq \alpha \leq 1$. Now, take any $0 \leq a, b \leq 1$. Then there are $0 \leq c, d \leq 1$ and $0 \leq \alpha \leq 1$ such that $a = c^\alpha$ and $b = d^{1-\alpha}$ (in fact, you can always take $\alpha = \frac{1}{2}$). Then, by identity (\ref{identity2}) from above,
\[
\varphi'(ab) - \varphi'(1) = \varphi'(c^\alpha d^{1-\alpha}) - \varphi'(1) = \alpha\varphi'(c) + (1-\alpha)\varphi'(d) - \varphi(1)
\]
But by identity (\ref{identity3}) from above,
\begin{itemize}
\item $\alpha\varphi'(c) = \varphi'(c^\alpha) - (1-\alpha)\varphi'(1) = \varphi'(a) - (1-\alpha)\varphi'(1)$
\item $(1-\alpha)\varphi'(d) = \varphi'(d^{1-\alpha}) - \alpha\varphi'(1) = \varphi'(b) - \alpha\varphi'(1)$
\end{itemize}
So
\[
\varphi'(ab) = \varphi'(a) - (1-\alpha)\varphi'(1) + \varphi'(b) - \alpha\varphi'(1)
\]
iff
\[
\varphi'(ab) = \varphi'(a) + \varphi'(b) - \varphi'(1)
\]
for all $0 \leq a, b \leq 1$. And this is the Cauchy functional equation for the logarithmic function. So $\varphi'(x) = m\log x + k$, for some constants $m, k$. Hence, $\varphi(x) = m(x\log x - x) + kx + c$, for some constant $c$. As we noted above, if $\psi$ is a strictly convex function and $\theta(x) = \psi(x) + kx + c$, then $\theta$ generates the same Bregman divergence as $\psi$. And thus, $\D$ is a positive linear transformation of $\GKL$, as required. \hfill $\Box$

\subsection{Proof of Theorem \ref{d-agg-wcap}} 

\noindent \emph{Proof of Theorem \ref{d-agg-wcap}. } By Lemma \ref{useful-lem}(i), $c^* = \Agg^{\{\alpha\}}_{\D_1}(c_1, \ldots, c_n) = \underset{c' \in C_\F}{\mathrm{arg\, min\ }} \sum^n_{k=1} \alpha_k \D(c', c_k)$ iff, for all $1 \leq j \leq n$,
\[
\varphi'(c^*(X_j)) = \sum^n_{k=1} \alpha_k \varphi'(c_k(X_j)).
\]
Then, by Corollary \ref{useful-col}(i), $c^\dag = \Fix_{\D_1}(\Agg^{\{\alpha\}}_{\D_1}(c_1, \ldots, c_n)) = \Fix_{\D_1}(c^*) = \underset{c' \in P_\F}{\mathrm{arg\, min\ }} \D(c', c^*)$ iff $c^\dag$ is coherent and there is a constant $K$ such that, for all $1 \leq j \leq n$,
\[
\varphi'(c^\dag(X_j)) - \varphi'(c^*(X_j)) = K
\]
iff $c^\dag$ is coherent and there is a constant $K$ such that, for all $1 \leq j \leq n$,
\[
\varphi'(c^\dag(X_j)) - \sum^n_{k=1} \alpha_k  \varphi'(c_k(X_j)) = K
\]
iff $c^\dag = \WCAP^{\{\alpha\}}_{\D}(c_1, \ldots, c_n) = \underset{c' \in P_\F}{\mathrm{arg\, min\ }} \sum^n_{k=1} \alpha_k \D(c', c_k)$, by Lemma \ref{useful-lem}(ii).\medskip

\noindent Furthermore, $c_k^\dag = \Fix_{\D_1}(c_k) =  \underset{c' \in P_\F}{\mathrm{arg\, min\ }} \D(c', c_k)$ iff $c^\dag$ is coherent and there is a constant $K$ such that, for all $1 \leq j \leq n$,
\[
\varphi'(c_k^\dag(X_j)) - \varphi'(c_k(X_j)) = K
\]
And $c^* = \Agg^{\{\alpha\}}_{\D_1}(c^\dag_1, \ldots, c^\dag_n) = \underset{c' \in C_\F}{\mathrm{arg\, min\ }} \sum^n_{k=1} \alpha_k \D(c', c^\dag_k)$ iff, for all $1 \leq j \leq n$,
%\[
%\sum^n_{k=1} \alpha_k  \left (\varphi'(c^*(X_j)) - \varphi'(c^\dag_k(X_j)) \right ) = 0
%\]
%iff, for all $1 \leq j \leq n$,
\[
\varphi'(c^*(X_j)) = \sum^n_{k=1} \alpha_k  \varphi'(c^\dag_k(X_j))
\]
Now, 
\begin{eqnarray*}
& &\varphi'(c^*(X_j)) - \sum^n_{k=1} \alpha_k  \varphi'(c_k(X_j)) \\
& = & \sum^n_{k=1} \alpha_k  \varphi'(c^\dag_k(X_j)) - \sum^n_{k=1} \alpha_k  \varphi'(c_k(X_j)) \\
& = & \sum^n_{k=1} \alpha_k  \left  (\varphi'(c^\dag_k(X_j)) - \varphi'(c_k(X_j)) \right )\\
& = & K
\end{eqnarray*}
Thus, $c^* = \Agg^{\{\alpha\}}_{\D_1}(\Fix_{\D_1}(c_1), \ldots, \Fix_{\D_1}(c_n))$ iff $c^*$ is coherent and there is a constant $K$ such that, for all $1 \leq j \leq n$,
\[
\varphi'(c^*(X_j)) - \sum^n_{k=1} \alpha_k  \varphi'(c_k(X_j)) = K
\]
iff $c^* = \WCAP^{\{\alpha\}}_{\D_1}(c_1, \ldots, c_n) = \underset{c' \in P_\F}{\mathrm{arg\, min\ }} \sum^n_{k=1} \alpha_k \D(c', c_k)$, by Lemma \ref{useful-lem}(ii).  \hfill $\Box$

\subsection{Proof of Theorem \ref{sed-gkl-2-necessary} }

\noindent \emph{Proof of Theorem \ref{sed-gkl-2-necessary}(i) } By Lemma \ref{useful-lem-2}(ii), $c^* = \WCAP^{\{\alpha\}}_{\D_2}(c_1, \ldots, c_n) = \underset{c' \in P_\F}{\mathrm{arg\, min}\ } \sum^n_{k=1} \alpha_k \D(c_k, c')$ iff $c^*$ is coherent and there is a constant $K$ such that 
\[
(c^*(X_j) - \sum^n_{k=1}\alpha_k c_k(X_j))\varphi''(c^*(X_j)) = K
\] 
for all $1 \leq j \leq m$. By Corollary \ref{useful-col}(ii), $c^* = \Fix_{\D_2}(\LP^{\{\alpha\}}(c_1, \ldots, c_n)) =  \Fix_{\D_2}(\sum^n_{k=1} \alpha_k c_k) = \underset{c' \in P_\F}{\mathrm{arg\, min}\ } \D(\sum^n_{k=1} \alpha_k c_k, c')$ iff $c^*$ is coherent and there is a constant $K$ such that 
\[
(c^*(X_j) - \sum^n_{k=1}\alpha_k c_k(X_j))\varphi''(c^*(X_j)) = K
\]
for all $1 \leq j \leq m$. Thus, $\Fix_{\D_2} \circ \LP = \WCAP_{\D_2}$, as required. \hfill $\Box$\bigskip

\noindent \emph{Proof of Theorem \ref{sed-gkl-2-necessary}(ii) } If $c_1, \ldots, c_n$ are coherent, then
\[
\Fix_{\D_2}(\LP^{\{\alpha\}}(c_1, \ldots, c_n)) = \LP^{\{\alpha\}}(c_1, \ldots, c_n) = \LP^{\{\alpha\}}(\Fix_{\D_2}(c_1), \ldots, \Fix_{\D_2}(c_n))
\]
% By Lemma \ref{useful-lem-2}(ii), $c^* = \WCAP^{\{\alpha\}}_{\D_2}(c_1, \ldots, c_n) = \underset{c' \in P_\F}{\mathrm{arg\, min}\ } \sum^n_{k=1} \alpha_k \D(c_k, c')$ iff  $c^*$ is coherent and there is a constant $K$ such that 
%\[
%(c^*(X_j) - \sum^n_{k=1}\alpha_k c_k(X_j))\varphi''(c^*(X_j)) = K
%\]
%for all $1 \leq j \leq m$. Now, suppose $c_1, \ldots, c_n$ are all coherent. And let $c^*(X_j) = \sum^n_{k=1} \alpha_k c_k(X_j)$. Then $c^*$ is also coherent and
%\[
%(c^*(X_j) - \sum^n_{k=1}\alpha_k c_k(X_j))\varphi''(c^*(X_j)) = 0%
%\]
as required. \hfill $\Box$ \bigskip

\noindent \emph{Proof of Theorem \ref{sed-gkl-2-necessary}(iii) } Suppose $c_1, c_2$ are defined on the partition $X_1$, $X_2$, $X_3$. By Lemma \ref{useful-lem}(i), $c_{\alpha}^* = \WCAP^{\{\alpha\}}_{\D_2}(c_1, c_2) = \underset{c' \in P_\F}{\mathrm{arg\, min}\ } \alpha \D(c_1, c') + (1-\alpha)\D(c_2, c')$ iff $c_{\alpha}^*$ is coherent and
\begin{eqnarray*}
& & (c_{\alpha}^*(X_i) - (\alpha c_1(X_i) + (1-\alpha) c_2(X_i)))\varphi''(c_{\alpha}^*(X_i)) \\
& = & (c_{\alpha}^*(X_j) - (\alpha c_1(X_j) + (1-\alpha) c_2(X_j)))\varphi''(c_{\alpha}^*(X_j))
\end{eqnarray*}
for all $1 \leq i, j \leq m$. By Corollary \ref{useful-col}(ii), $c_k^* = \Fix_{\D_2}(c_k) = \underset{c' \in P_\F}{\mathrm{arg\, min}\ } \D(c_k, c')$ iff $c_k^*$ is coherent and
\[
(c_k^*(X_i) - c_k(X_i))\varphi''(c_k^*(X_i)) = (c_k^*(X_j) - c_k(X_j))\varphi''(c_k^*(X_j))
\]
for all $1 \leq i, j \leq m$. Since $\WCAP^{\{\alpha\}}_{\D_2} = \LP \circ \Fix_{\D_2}$,
\begin{eqnarray*}
& & ((\alpha c^*_1(X_i) + (1-\alpha) c^*_2(X_i)) - (\alpha c_1(X_i) + (1-\alpha) c_2(X_i)))\varphi''(\alpha c^*_1(X_i) + (1-\alpha) c^*_2(X_i)) \\
& = & ((\alpha c^*_1(X_j) + (1-\alpha) c^*_2(X_j)) - (\alpha c_1(X_j) + (1-\alpha) c_2(X_j)))\varphi''(\alpha c^*_1(X_j) + (1-\alpha) c^*_2(X_j))
\end{eqnarray*}
for all $1 \leq i, j \leq m$, and this holds iff
\begin{eqnarray*}
& & (\alpha ( c^*_1(X_i) - c_1(X_i)) + (1-\alpha)(c^*_2(X_i) - c_2(X_i)))\varphi''(\alpha c^*_1(X_i) + (1-\alpha) c^*_2(X_i)) \\
& = & (\alpha ( c^*_1(X_j) - c_1(X_j)) + (1-\alpha)(c^*_2(X_j) - c_2(X_j)))\varphi''(\alpha c^*_1(X_j) + (1-\alpha) c^*_2(X_j))
\end{eqnarray*}
for all $1 \leq i, j \leq m$. Now, for all coherent credence functions $c$, $c^* = c$. So, if $\alpha > 0$, then
\begin{eqnarray*}
( c^*_1(X_1) - c_1(X_1))\varphi''(\alpha c^*_1(X_1) + (1-\alpha) c(X_1)) \\
= ( c^*_1(X_2) - c_1(X_2)) \varphi''(\alpha c^*_1(X_2) + (1-\alpha) c(X_2))
\end{eqnarray*}
Now, given any $0 < x, y < 1$, we can find $0 \leq \alpha \leq 1$ and $0 \leq a, b \leq 1$ such that \begin{itemize}
\item $\alpha c^*_1(X_1) + (1-\alpha)a = x$
\item $\alpha c^*_1(X_1) + (1-\alpha)b = y$. 
\end{itemize}
Then let $c$, $c'$ be the following  coherent credence functions:
\begin{itemize}
\item $c(X_1) = a$, $c(X_2) = 0$, $c(X_3) = 1-a$
\item $c'(X_1) = b$, $c'(X_2) = 0$, $c'(X_3) = 1-b$.
\end{itemize}
Then
\begin{eqnarray*}
& & ( c^*_1(X_1) - c_1(X_1))\varphi''(x) \\
& = & ( c^*_1(X_1) - c_1(X_1))\varphi''(\alpha c^*_1(X_1) + (1-\alpha)a) \\
& = & ( c^*_1(X_1) - c_1(X_1))\varphi''(\alpha c^*_1(X_1) + (1-\alpha)c(X_1)) \\
& = & ( c^*_1(X_2) - c_1(X_2))\varphi''(\alpha c^*_1(X_2) + (1-\alpha)c(X_2)) \\
& = & ( c^*_1(X_2) - c_1(X_2))\varphi''(\alpha c^*_1(X_2) + (1-\alpha)\cdot 0) \\
& = & ( c^*_1(X_2) - c_1(X_2))\varphi''(\alpha c^*_1(X_2) + (1-\alpha)c'(X_2)) \\
& = & ( c^*_1(X_1) - c_1(X_1))\varphi''(\alpha c^*_1(X_1) + (1-\alpha)c'(X_1)) \\
& = & ( c^*_1(X_1) - c_1(X_1))\varphi''(\alpha c^*_1(X_1) + (1-\alpha)b) \\
& = & ( c^*_1(X_1) - c_1(X_1)) \varphi''(y)
\end{eqnarray*}
Thus, $\varphi''(x) = \varphi''(y)$, for all $0 < x, y < 1$. Thus, for all $0 \leq x \leq 1$, $\varphi''(x) = m$, so $\varphi'(x) = mx + k$, so $\varphi(x) = m'x^2 + kx + c$, for some constants $m, m', k, c$. Thus $\D$ is a positive linear transformation of $\SED$, as required. \hfill $\Box$

\subsection{Proof of Theorems \ref{agg-unique-1} and \ref{agg-unique-2}}

\emph{Proof of Theorem \ref{agg-unique-1}(i) } By Lemma \ref{useful-lem}(i), $c^* = \Agg^{\{\alpha\}}_{\D_1}(c_1, \ldots, c_n) = \underset{c' \in C_\F}{\mathrm{arg\, min}\ } \sum^n_{k=1} \alpha_k \D(c', c_k)$ iff, for all $1 \leq j \leq m$,
\[
\varphi'(c^*(X_j)) = \sum^n_{k=1} \alpha_k \varphi'(c_k(X_j))
\]
And of course $c^* = \LP^{\{\alpha\}}(c_1, \ldots, c_n)$ iff,  for all $1 \leq j \leq m$,
\[
c^*(X_j) = \sum^n_{k=1} \alpha_k c_k(X_j)
\]
Thus, $\Agg_{\D_1} = \LP$ iff, for any $\alpha_1, \ldots, \alpha_n$, and $c_1, \ldots, c_n$,
\[
\varphi'\left (\sum^n_{k=1} \alpha_k c_k(X_j)\right ) = \sum^n_{k=1} \alpha_k \varphi'(c_k(X_j))
\]
iff, for any $0 \leq x, y, \leq 1$, and $0 \leq \alpha \leq 1$,
\[
\varphi'(\alpha x + (1-\alpha)y) = \alpha\varphi'(x) + (1-\alpha)\varphi'(y)
\]
And thus, $\varphi'(x) = kx + c$, for some constants $k$, $c$. From this point, the proof proceeds in the same fashion as the proof of Theorem \ref{sed-gkl-1-necessary}(i).\hfill $\Box$\medskip

\noindent \emph{Proof of Theorem \ref{agg-unique-1}(ii) } Again, by Lemma \ref{useful-lem}(i), $c^* = \Agg^{\{\alpha\}}_{\D_1}(c_1, \ldots, c_n) = \underset{c' \in C_\F}{\mathrm{arg\, min}\ } \sum^n_{k=1} \alpha_k \D(c', c_k)$ iff, for all $1 \leq j \leq m$,
\[
\varphi'(c^*(X_j)) = \sum^n_{k=1} \alpha_k \varphi'(c_k(X_j))
\]
And of course $c^* = \GP_-^{\{\alpha\}}(c_1, \ldots, c_n)$ iff,  for all $1 \leq j \leq m$,
\[
c^*(X_j) = \prod^n_{k=1} c_k(X_j)^{\alpha_k}
\]
Thus, $\Agg_{\D_1} = \GP_-$ iff, for any $\alpha_1, \ldots, \alpha_n$, and $c_1, \ldots, c_n$,
\[
\varphi'\left (\prod^n_{k=1} c_k(X_j)^{\alpha_k}\right ) = \sum^n_{k=1} \alpha_k \varphi'(c_k(X_j))
\]
iff, for any $0 \leq x, y \leq 1$, and $0 \leq \alpha \leq 1$,
\[
\varphi'(x^\alpha y^{1-\alpha}) = \alpha\varphi'(x) + (1-\alpha)\varphi'(y)
\]
And thus, $\varphi'(x) = m\log x + k$, for some constants $m$, $k$. From this point on, the proof proceeds in the same fashion as the proof of Theorem \ref{sed-gkl-1-necessary}(ii).\hfill $\Box$\medskip

\noindent \emph{Proof of Theorem \ref{agg-unique-2}(i) } By Lemma \ref{useful-lem-2}(i), $c^* = \Agg^{\{\alpha\}}_{\D_2}(c_1, \ldots, c_n) = \underset{c' \in C_\F}{\mathrm{arg\, min}\ } \sum^n_{k=1} \alpha_k \D(c_k, c')$ iff, for all $1 \leq j \leq m$,
\[
\left (c^*(X_j) - \sum^n_{k=1} \alpha_k c_k(X_j) \right ) \varphi''(c^*(X_j)) =0
\]
And of course $c^* = \LP^{\{\alpha\}}(c_1, \ldots, c_n)$ iff,  for all $1 \leq j \leq m$,
\[
c^*(X_j) = \sum^n_{k=1} \alpha_k c_k(X_j)
\]
Thus, $\Agg_{\D_2} = \LP$ iff 
\[
\left ( \sum^n_{k=1} \alpha_k c_k(X_j) - \sum^n_{k=1} \alpha_k c_k(X_j) \right ) \varphi''(c^*(X_j)) =0
\]
And that is true for any $\D$. \hfill $\Box$

\bibliographystyle{/Users/rp3959/Dropbox/apa-good}
\bibliography{/Users/rp3959/Dropbox/bibliography}

\end{document}